\documentclass[useAMS,usenatbib]{mn2e}
\usepackage{graphicx}
\usepackage{txfonts
}
\usepackage{color}

\title[ The Large-Scale Filament Feeding MACSJ0717.5+3745]{A Weak-Lensing Mass Reconstruction of the Large-Scale Filament Feeding the Massive Galaxy Cluster MACSJ0717.5+3745}
\author[M. Jauzac, E. Jullo, J.-P. Kneib, H. Ebeling, A. Leauthaud, C. J. Ma, M. Limousin, R. Massey, J. Richard]
{Mathilde Jauzac,$^{1,2}$\thanks{E-mail: mathilde.jauzac@gmail.com (MJ)}
Eric Jullo,$^{1,3}$ Jean-Paul Kneib,$^1$ Harald Ebeling,$^4$ Alexie Leauthaud,$^{5}$ 
\newauthor
 Cheng-Jiun Ma,$^4$ Marceau Limousin,$^{1,6}$ Richard Massey,$^{7}$ Johan Richard$^8$\\
\\
\\
$^{1}$Laboratoire d'Astrophysique de Marseille - LAM, Universit\'e d'Aix-Marseille $\&$ CNRS, UMR7326, 38 rue F. Joliot-Curie, 13388 Marseille Cedex 13, France\\
$^{2}$Astrophysics and Cosmology Research Unit, School of Mathematical Sciences, University of KwaZulu-Natal, Durban 4041, South Africa\\
$^{3}$Jet Propulsion Laboratory, California Institute of Technology, Pasadena, CA 91109, USA\\
$^{4}$Institute for Astronomy, University of Hawaii, 2680 Woodlawn Drive, Honolulu, Hawaii 96822, USA\\
$^{5}$Kavli Institute for the Physics and Mathematics of the Universe, Todai Institutes for Advanced Study, the University of Tokyo, Kashiwa, Japan 277-8583\\
(Kavli IPMU, WPI)\\
$^{6}$Dark Cosmology Centre, Niels Bohr Institute, University of Copenhagen, Juliane Maries Vej 30, DK-2100 Copenhagen, Denmark\\
$^{7}$Institute for Computational Cosmology, Durham University, South Road, Durham DH1 3LE, U.K.\\
$^{8}$CRAL, Observatoire de Lyon, Universit\'e Lyon 1, 9 Avenue Ch. Andr\'e, 69561 Saint Genis Laval Cedex, France}
\begin{document}

\date{Accepted 2012 August 20. Received 2012 August 15; in original form: 2012 April 27}

\pagerange{\pageref{firstpage}--\pageref{lastpage}} \pubyear{XXXX}

\maketitle

\label{firstpage}

\begin{abstract}
We report the first weak-lensing detection of a large-scale filament funneling matter onto the core of the massive galaxy cluster MACSJ0717.5+3745. 

Our analysis is based on a mosaic of 18 multi-passband images obtained with the Advanced Camera for Surveys aboard the Hubble Space Telescope, covering an area of $\sim 10 \times 20$ arcmin$^{2}$. We use a weak-lensing pipeline developed for the COSMOS survey, modified for the analysis of galaxy clusters, to produce a weak-lensing catalogue. A mass map is then computed by applying a weak-gravitational-lensing multi-scale reconstruction technique designed to describe irregular mass distributions such as the one investigated here. We test the resulting mass map by comparing the mass distribution inferred for the cluster core with the one derived from strong-lensing constraints and find excellent agreement.

Our analysis detects the MACSJ0717.5+3745 filament within the 3~sigma detection contour of the lensing mass reconstruction, and underlines the importance of filaments for theoretical and numerical models of the mass distribution in the Cosmic Web. We measure the filament's projected length as $\sim$ 4.5 $h_{74}^{-1}$ Mpc, and its mean density as $(2.92 \pm 0.66)\times10^{8}~h_{74}$ M$_{\odot}$ kpc$^{-2}$. 
Combined with the redshift distribution of galaxies obtained after an extensive spectroscopic follow-up in the area, we can rule out any projection effect resulting from the chance alignment on the sky of unrelated galaxy group-scale structures. Assuming plausible constraints concerning the structure's geometry based on its galaxy velocity field, we construct a 3D model of the large-scale filament. Within this framework, we derive the three-dimensional length of the filament to be 18~$h_{74}^{-1}$ Mpc.  The filament's deprojected density in terms of the critical density of the Universe is measured as $(206 \pm 46)\times \rho_{\rm crit}$, a value that lies at the very high end of the range predicted by numerical simulations. 
Finally, we study the distribution of stellar mass in the field of MACSJ0717.5+3749 and, adopting a mean mass-to-light ratio $\langle M_{\ast}/L_{K}\rangle$ of $0.73 \pm 0.22$ and assuming a Chabrier Initial-Mass Function, measure a stellar mass fraction along the filament of $(0.9 \pm 0.2)$\%, consistent with previous measurements in the vicinity of massive clusters.
\end{abstract}


%______________________________________________________
% SECTION 1 :  INTRODUCTION
%______________________________________________________
\section{Introduction}
In a Universe dominated by Cold Dark Matter (CDM), such as the one parameterised by the $\Lambda$CDM concordance cosmology, hierarchical structure formation causes massive galaxy clusters to form through a series of successive mergers of smaller clusters and groups of galaxies, as well as through continuous accretion of  surrounding matter. % channeled through filaments. 
N-body simulations of the dark-matter distribution on very large scales \citep{bond96,YS96,aragoncalvo07,hahn07} predict that these processes of merging and accretion occur along preferred directions, i.e., highly anisotropically. The result is the ``cosmic web'' \citep{bond96}, a spatially highly correlated structure of interconnected filaments and vertices marked by massive galaxy clusters. 
Abundant observational support for this picture has been provided by large-scale galaxy redshift surveys \citep[e.g.,][] {GH89, york00,colless01} showing voids surrounded and connected by filaments and sheets of galaxies. 

A variety of methods have been developed to detect filaments in surveys, among them a ``friends of friends'' algorithm \citep[FOF,][]{HG82} combined with ``Shapefinders'' statistics \citep{SJ03}; the ``Skeleton'' algorithm \citep{novikov06,sousbie06a}; a two-dimensional technique developed by \citet{moody83}; and the Smoothed Hessian Major Axis Filament Finder \citep[SHMAFF,][]{bond10}.

Although ubiquitous in large-scale galaxy surveys, filaments have proven hard to characterise physically, owing to their low density and the fact that the best observational candidates often turn out to be not primordial in nature but the result of recent cluster mergers. Specifically, attempts to study the warm-hot intergalactic medium \citep[WHIM,][]{CO99}, resulting from the expected gravitational heating of the intergalactic medium in filaments, remain largely inconclusive because it is hard to ascertain for filaments near cluster whether spectral X-ray features originate from the filament or from past or ongoing clusters mergers \citep{kaastra06,rasmussen07,galeazzi09,williams10}. Some detections appear robust as they have been repeatedly confirmed \citep{fang02,fang07,williams07} but are based on just one X-ray line. An alternative observational method is based on a search for filamentary overdensities of galaxies relative to the background \citep{PD04,ebeling04}. When conducted in 3D, i.e., including spectroscopic galaxy redshifts, this method is well suited to detecting filament candidates. It does, however, not allow the determination of key physical properties unless it is supplemented by follow-up studies targeting the presumed WHIM and dark matter which are expected to constitute the vast majority of the mass of large-scale filaments. By contrast, weak gravitational lensing offers the tantalising possibility of detecting directly the total mass content of filaments \citep{mead10}, since the weak-lensing signal arises from luminous and dark matter alike, regardless of its dynamical state. 

%______________________________________________________
% Fig 1 
%______________________________________________________
\begin{figure}%[!h]
\includegraphics[width=84mm]{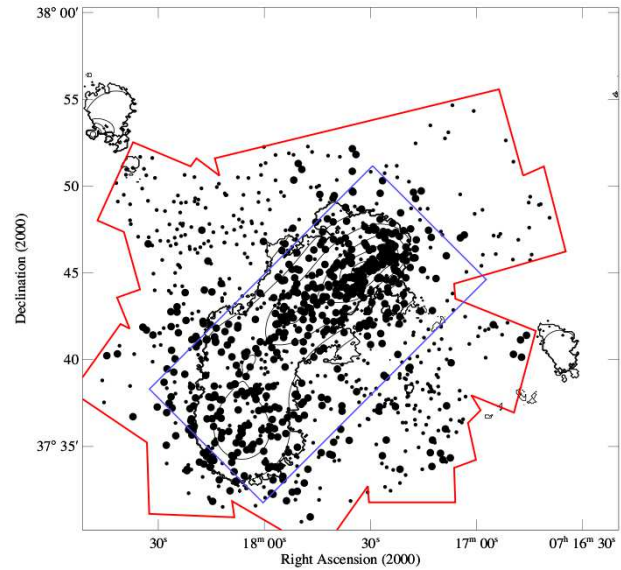}
\caption{Area of our spectroscopic survey of MACSJ0717.5+3745. Outlined in red is the region covered by our Keck/DEIMOS masks; outlined in blue is the area observed with HST/ACS. Small circles correspond to objects for which redshifts were obtained; large filled circles mark cluster members. The black contours show the projected galaxy density (see Ma et al.\ 2008).}
\label{fig:area0717}
\end{figure}
%%%%%%%%%%%%%%%%%%%%%%%%%%%%%%%%%%%%

Previous weak gravitational lensing studies of binary clusters found tentative evidence of filaments, but did not result in clear detections. One of the first efforts was made by \citet{clowe98} who reported the detection of a filament apparently extending from the distant cluster RX\,J1716+67 ($z=0.81$), using images obtained with the Keck 10m telescope and University of Hawaii (UH) 2.2m telescope. This filamentary structure would relate two distinct sub clusters detected on the mass and light maps. The detection was not confirmed though. Almost at the same time \citet{kaiser98} conducted a weak lensing study of the supercluster MS0302+17 with the UH8K CCD camera on the Canada France Hawaii Telescope (CFHT). Their claimed detection of a filament in the field was, however, questioned on the grounds that the putative filament overlapped with both a foreground structure as well as with gaps between CCD chips. Indeed, \citet{gavazzi04} showed the detection to have been spurious by means of a second study of MS0302+17 using the CFHT12K camera. A weak gravitational lensing analysis with MPG/ESO Wide Field Imager conducted by  \citet{gray02}  claimed the detection of a filament in the triple cluster A901/902. The candidate filament appeared to connect two of the clusters and was detected in both the galaxy distribution and in the weak-lensing mass map. However, this detection too was of low significance and coincided partly with a gap between two chips of the camera. As in the case of MS0302+17, a re-analysis of the A901/A902 complex using high-quality HST/ACS images by \citet{heymans08} failed to detect the filament and led the authors to conclude that the earlier detection was caused by residual PSF systematics and limitation of the KS93 mass reconstruction used in the study by \citet{gray02}. A further detection of a filament candidate was reported by \citet{dietrich05} based on a weak gravitational lensing analysis of the close double cluster A222/A223. However, as in other similar cases, the proximity of the two clusters connected by the putative filament raises the possibility of the latter being a merger remnant rather than primordial in nature.

 %__________________________________________________________________
% TABLE 1 
 %__________________________________________________________________
\begin{table*}%[!t]
\caption{Overview of the HST/ACS observations of MACSJ0717.5+3745.
*: Cluster core; observed through F555W, rather than F606W filter.}
\label{tab:acs_obs}
\begin{center}
\begin{tabular}[h]{|ccccccc}
\hline\\[-5pt]
                     &   & \multicolumn{2}{c}{F606W} & \multicolumn{2}{c}{F814W} &  \\ \cline{3-4} \cline{5-6} \\[-5pt]
R.A.\ (J2000) & Dec (J2000) & Date & Exposure Time (s) & Date & Exposure Time (s) & Programme  ID \\
 \hline\\[-5pt]
07 17 32.93 & +37 45 05.4 & \,\,\,2004-04-02* & \,\,\,4470* & 2004-04-02 & 4560 & \,\,\,\,9722 \\
07 17 31.81 & +37 49 20.6 & 2005-02-08 & 1980 & 2005-02-08 & 4020 & 10420 \\
07 17 20.38 & +37 47 07.5 & 2005-01-27 & 1980 & 2005-01-27 & 4020 & 10420 \\
07 17 08.95 & +37 44 54.3 & 2005-01-27 & 1980 & 2005-01-27 & 4020 & 10420 \\
07 17 43.23 & +37 47 03.1 & 2005-01-30 & 1980 & 2005-01-30 & 4020 & 10420 \\
07 17 20.18 & +37 42 38.8 & 2005-02-01 & 1980 & 2005-02-01 & 4020 & 10420 \\
07 17 54.26 & +37 44 49.3 & 2005-01-27 & 1980 & 2005-01-27 & 4020 & 10420 \\
07 17 42.82 & +37 42 36.3 & 2005-01-24 & 1980 & 2005-01-25 & 4020 & 10420 \\
07 17 31.39 & +37 40 23.3 & 2005-02-01 & 1980 & 2005-02-01 & 4020 & 10420 \\
07 18 05.46 & +37 42 33.6 & 2005-02-04 & 1980 & 2005-02-04 & 4020 & 10420 \\
07 17 54.02 & +37 40 20.6 & 2005-02-04 & 1980 & 2005-02-04 & 4020 & 10420 \\
07 17 42.79 & +37 38 05.7 & 2005-02-05 & 1980 & 2005-02-05 & 4020 & 10420 \\
07 18 16.65 & +37 40 17.7 & 2005-02-05 & 1980 & 2005-02-05 & 4020 & 10420 \\
07 18 05.22 & +37 38 04.9 & 2005-02-05 & 1980 & 2005-02-05 & 4020 & 10420 \\
07 17 53.79 & +37 35 52.0 & 2005-02-05 & 1980 & 2005-02-05 & 4020 & 10420 \\
07 18 27.84 & +37 38 01.9 & 2005-02-08 & 1980 & 2005-02-08 & 4020 & 10420 \\
07 18 16.40 & +37 35 49.1 & 2005-02-08 & 1980 & 2005-02-08 & 4020 & 10420 \\
07 18 04.97 & +37 33 36.2 & 2005-02-09 & 1980 & 2005-02-09 & 4020 & 10420 \\
\hline\\[-5pt]
\end{tabular}\\
\end{center}
\end{table*}
%%%%%%%%%%%%%%%%%%%%%%%%%%%%%%%%%%%%%%%%%%%

In this paper we describe the first weak gravitational analysis of the very massive cluster MACSJ0717.5+3745 \citep[$z=0.55$;][]{edge03,ebeling04,ebeling07,ma08,ma09}. Optical and X-ray analyses of the system \citep{ebeling04, ma08,ma09} find compelling evidence of a filamentary structure extending toward the South-East of the cluster core. Using weak-lensing data to reconstruct the mass distribution in and around MACSJ0717.5+3745, we directly detect the reported filamentary structure in the field of MACSJ0717.5+3745.

The paper is organized as follows. After an overview of the observational data in Section 2, we discuss the gravitational lensing data in hand in Section 3. The modeling of the mass using a multi-scale approach is described in Section 4. Results are discussed in Section 5, and we present our conclusions in Section 6.
\\
\\
All our results use the $\Lambda$CDM concordance cosmology with $\Omega_{\rm M}$ = 0.3, $\Omega_{\Lambda}$ = 0.7, and a Hubble constant $H_{0}$ = 74 km s$^{-1}$ Mpc$^{-1}$, hence 1" corresponds to 6.065~kpc at the redshift of the cluster. Magnitudes are quoted in the AB system.

%______________________________________________________
% SECTION 2 :  OBSERVATIONS
%______________________________________________________
\section{Observations}
The MAssive Cluster Survey \citep[MACS,][]{ebeling01} was the first cluster survey to search exclusively for very massive clusters at moderate to high redshift. Covering over 20,000 deg$^2$ and using dedicated optical follow-up observations to identify faint X-ray sources detected in the ROSAT All-Sky Survey, MACS compiled a sample of over 120 very X-ray luminous clusters at $z>0.3$, thereby more than tripling the number of such systems previously known. The high-redshift MACS subsample \citep{ebeling07} comprises 12 clusters a  $z>0.5$. MACSJ0717.5+3745 is one of them. All 12 were observed with the ACIS-I imaging spectrograph onboard the Chandra X-ray Observatory. Moderately deep optical images covering $30\times 27$ arcmin$^2$ were obtained in five passbands (B, V, R, I, z$^\prime$) with the SuprimeCam wide-field imager on the Subaru 8.2m Telescope, and supplemented with u-band imaging obtained with MegaCam on the Canada France Hawaii Telescope (CFHT). Finally, the cores of all clusters in this MACS subsample were observed with the Advanced Camera for Surveys (ACS) onboard HST in two bands, F555W $\&$ F814W, for 4.5ks in both bands, as part of programmes GO-09722 and GO-11560 (PI Ebeling).

%______________________________________________________
%  2.1 : Imaging with HST
%______________________________________________________
\subsection{HST/ACS Wide-Field Imaging}
A mosaic of images of MACSJ0717.5+3745 and the filamentary structure to the South-East was obtained between January 24 and February 9, 2005, with the ACS aboard HST (GO-10420, PI Ebeling). The $3\times 6$ mosaic consists of images in the F606W and F814W filters, observed for roughly 2.0 ks and 4.0 ks respectively (1 $\&$ 2 HST orbits). Only 17 of the 18 tiles of the mosaic were covered though, since the core of the cluster had been observed already (see  Tab.~\ref{tab:acs_obs} for more details). 

Charge Transfer Inefficiency (CTI), due to radiation damage of the ACS CCDs above the Earth's atmosphere, creates spurious trails behind objects in HST/ACS images. Since CTI affects galaxy photometry, astrometry, and shape measurements, correcting the effect is critical for weak-lensing studies. We apply the algorithm proposed by  \citet{massey10} which operates on the raw data and returns individual electrons to the pixels from which they were erroneously dragged during readout. Image registration, geometric distortion corrections, sky subtraction, cosmic ray rejection, and the final combination of the dithered images are then performed using the standard MULTIDRIZZLE routines \citep{koekemoer02}. MULTIDRIZZLE parameters are set to values optimised for precise galaxy shape measurement \citep{rhodes07}, and output images created with a 0.03" pixel grid, compared to the native ACS pixel scale of 0.05".

%______________________________________________________
% Table 2 
%______________________________________________________
\begin{table}%[!t]
\caption{Overview of groundbased imaging observations of MACSJ0717.5+3745.}
\label{tab:subarucfht_obs}
\begin{center}
\begin{tabular}[h]{lccccc@{\hspace{5mm}}ccc}
\hline\\[-5pt]
& \multicolumn{5}{c}{Subaru} & \multicolumn{3}{c}{CFHT} \\ \cline{2-6}\cline{7-9}\\[-5pt]
 & B & V & $R_{C}$ & $I_{C}$ & z' & u* & J & $K_{S}$ \\
 \hline\\[-5pt]
Exposure (hr) & 0.4 & 0.6 & 0.8 & 0.4 & 0.5 & 1.9 & 1.8 & 1.7\\
Seeing (arcsec) & 0.8 & 0.7 & 1.0 & 0.8 & 0.6 & 1.0 & 0.9 & 0.7 \\
\hline\\[-5pt]
\end{tabular}\\
\end{center}
%\label{default}
\end{table}%
%%%%%%%%%%%%%%%%%%%%%%%%%%%%%%%%%%%%

%______________________________________________________
%  2.2 : Imaging with SUBARU
%______________________________________________________
\subsection{Groundbased Imaging}
MACSJ0717.5+3745 was observed in the B, V, $R_{c}$, $I_{c}$ and z$^{\prime}$ bands with the Suprime-Cam wide-field camera on the Subaru 8.2m telescope \citep{miyazaki02}. These observations are supplemented by images in the u* band obtained with the MegaPrime camera on the CFHT 3.6m telescope, as well as near-infrared imaging in the J and $K_{S}$ bands obtained with WIRcam on CFHT. Exposure times and seeing conditions for these observations are listed in Tab.~\ref{tab:subarucfht_obs} (see also C.-J.\ Ma, Ph.D.\ thesis). All data were reduced using standard techniques which were, however, adapted to deal with special characteristics of the Suprime-Cam and MegaPrime data; for more details see \cite{donovan07}.

The groundbased imaging data thus obtained are used primarily to compute photometric redshifts which allow the elimination of cluster members and foreground galaxies that would otherwise dilute the shear signal. To this end we use the object catalogue compiled by \cite{ma08} which we describe briefly in the following. Imaging data from the  passbands listed in Tab.~\ref{tab:subarucfht_obs} were seeing-matched using the technique described in \cite{kartaltepe08} in order to allow a robust estimate of the spectral energy distribution (SED) for all objects within the field of view. The object catalogue was then created using the SE\textsc{xtractor} photometry package \citep{BA96} in "dual mode" using the R-band image as the reference detection image. More details are given in \cite{ma08}.

%______________________________________________________
% Fig 2
%______________________________________________________
\begin{figure}%[!t]
\includegraphics[width=88mm]{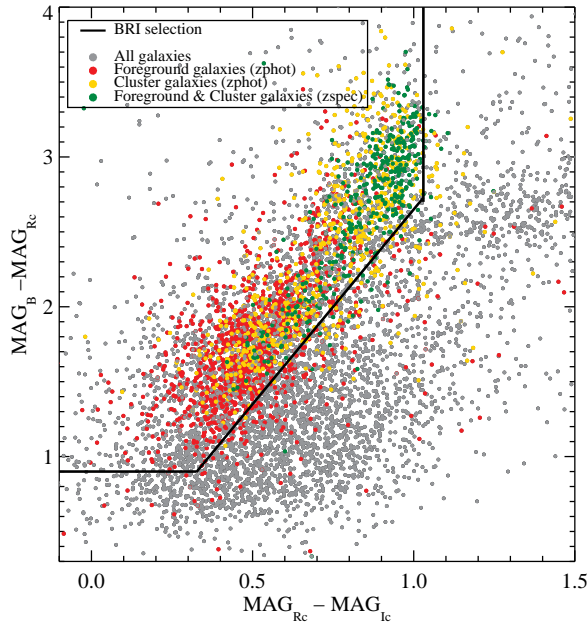}
\caption{Colour-colour diagram (B${-}$R vs R${-}$I) for objects within the HST/ACS mosaic of MACSJ0717.5+3745. Grey dots represent all objects in the study area. Unlensed galaxies diluting the shear signal are marked by different colours: galaxies spectroscopically confirmed as cluster members or foreground galaxies (green);  galaxies classified as foreground objects because of their photometric redshifts (red); and galaxies classified as cluster members via photometric redshifts (yellow). The solid black lines delineate the BRI colour-cut defined for this work to mitigate shear dilution by unlensed galaxies.}
\label{fig:colorcolor0717}
\end{figure}
%%%%%%%%%%%%%%%%%%%%%%%%%%%%%%%%%%%%

%______________________________________________________
% 2.3
%______________________________________________________
\subsection{Spectroscopic and Photometric Redshifts}
\label{sect:sec23}
Spectroscopic observations of MACSJ0717.5+3745 (including the full length of the filament) were conducted between 2000 and 2008, mainly with the DEIMOS spectrograph on the Keck-II 10m telescope on Mauna Kea, supplemented by observations of the cluster core region performed with the LRIS and GMOS spectrographs on Keck-I and Gemini-North, respectively. The DEIMOS instrument setup combined the 600ZD grating with the GC455 order-blocking filter and a central wavelength between 6300 and 7000~\AA; the exposure time per MOS (multi-object spectroscopy) mask was typically 3$\times$1800\,s. A total of 18 MOS masks were used in our DEIMOS observations; spectra of 1752 unique objects were obtained (65 of them with LRIS, and 48 with GMOS), yielding 1079 redshifts, 537 of them of cluster members.  Figure~\ref{fig:area0717} shows the area covered by our spectroscopic survey as well as the loci of the targeted galaxies. The data were reduced with the DEIMOS pipeline developed by the DEEP2 project.

Photometric redshifts for galaxies with $m_{\rm R_c}{<}24.0$ were computed using the adaptive SED-fitting code Le Phare \citep{arnouts99,ilbert06,ilbert09}. In addition to employing $\chi^2$ optimization during SED fitting, Le Phare adaptively adjusts the photometric zero points by using galaxies with spectroscopic redshifts as a training set. This approach reduces the fraction of catastrophic errors and also mitigates systematic trends in the differences between spectroscopic and photometric redshifts \citep{ilbert06}.

Further details, e.g.\ concerning the selection of targets for spectroscopy or the spectral templates used for the determination of photometric redshifts, are provided by \cite{ma08}. The full redshift catalogue as well as an analysis of cluster substructure and dynamics as revealed by radial velocities will be presented in Ebeling et al.\ (2012, in preparation). 

%______________________________________________________
% Fig 3
%______________________________________________________
\begin{figure}%[!t]
\includegraphics[width=88mm]{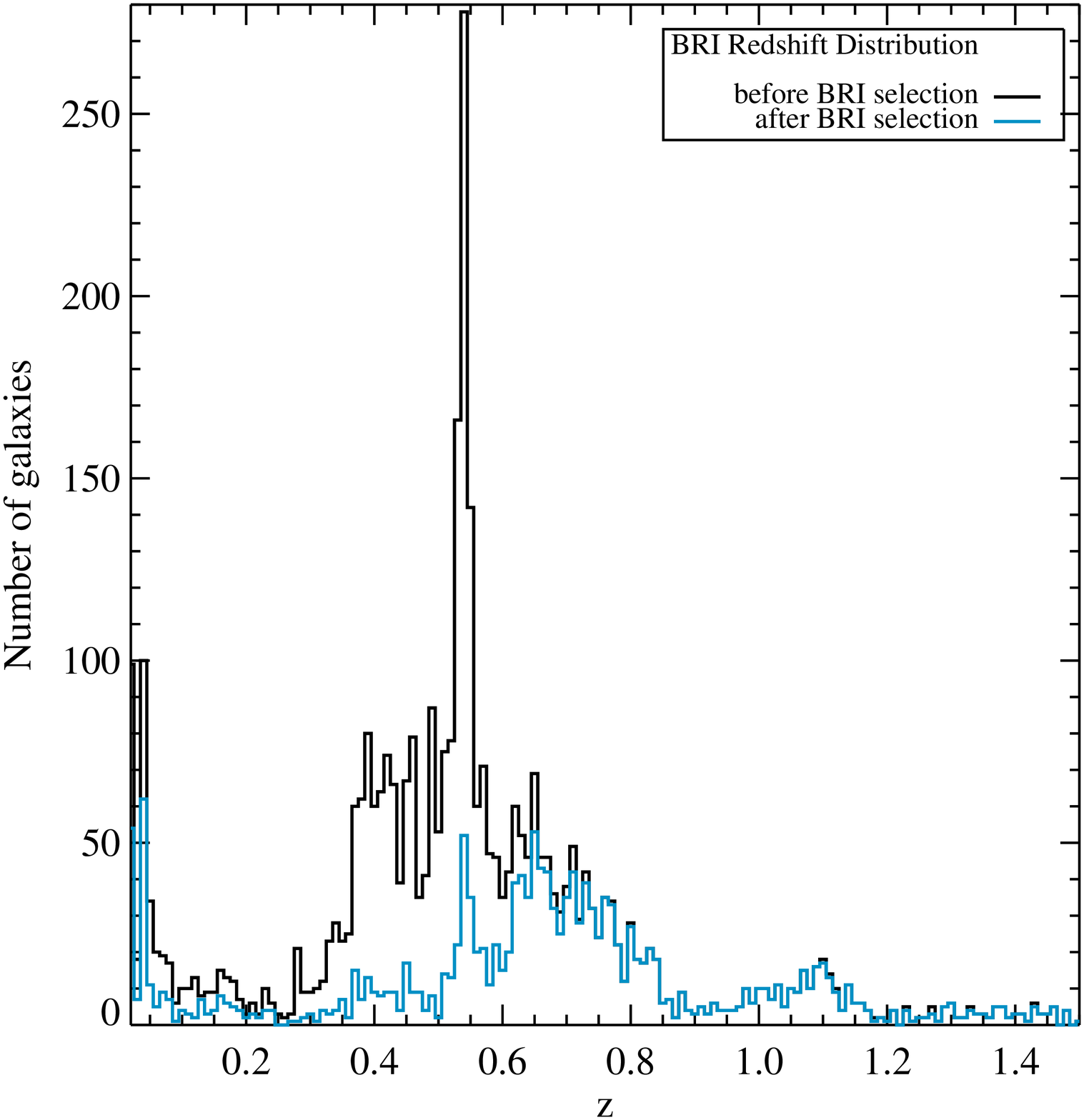}
\caption{Redshift distribution of all galaxies with B, R$_{c}$, and I$_{c}$ photometry from Subaru/SuprimeCam observations that have photometric or spectroscopic redshifts (black histogram). The cyan histogram shows the redshift distribution of galaxies classified as background objects using the BRI criterion illustrated in Fig.~\ref{fig:colorcolor0717}.}
\label{fig:Nzbri}
\end{figure}
%%%%%%%%%%%%%%%%%%%%%%%%%%%%%%%%%%%%

%______________________________________________________
% SECTION 3 : WGL ANALYSIS
%______________________________________________________
\section{Weak Gravitational Lensing Analysis}
\label{sect:sec3}
%______________________________________________________
% 3.1
%______________________________________________________
\subsection{The ACS catalogue}
Our weak-lensing analysis is based on shape measurements in the ACS/F814W band.
Following a method developed for the analysis of data obtained for the COSMOS survey and described in \cite{leauthaud07} (hereafter L07) we use the SE\textsc{xtractor} photometry package \citep{BA96} to detect sources in our ACS imaging data in a two-step process. Called the ``Hot-Cold" technique \citep[][L07]{rix04}, it consists of running SE\textsc{xtractor} twice: first with a configuration optimised for the detection of only the brightest objects (the ``cold" step), then a second time with a configuration optimised for the detection of the faint objects (the ``hot" step) that contain most of the lensing signal. The resulting object catalogue is then cleaned by removing spurious or duplicate detections using a semi-automatic algorithm that defines polygonal masks around stars or saturated pixels. 

Star-galaxy classification is performed by examining the distribution of objects in the magnitude (MAG$\_$AUTO) vs peak surface-brightness (MU$\_$MAX) plane. This diagram allows us to separate three classes of objects: galaxies, stars, and any remaining spurious detections (i.e., artifacts, hot pixels and residual cosmic rays).
Finally, the drizzling process introduces pattern-dependent correlations between neighbouring pixels which artificially reduces the noise level of co-added drizzled images. We apply the remedy used by L07 by simply scaling up the noise level in each pixel by the same constant $F_{A} \approx 0.316$, defined by \cite{casertano00}.

%______________________________________________________
% Fig 4
%______________________________________________________
\begin{figure}%[!t]
\includegraphics[width=88mm]{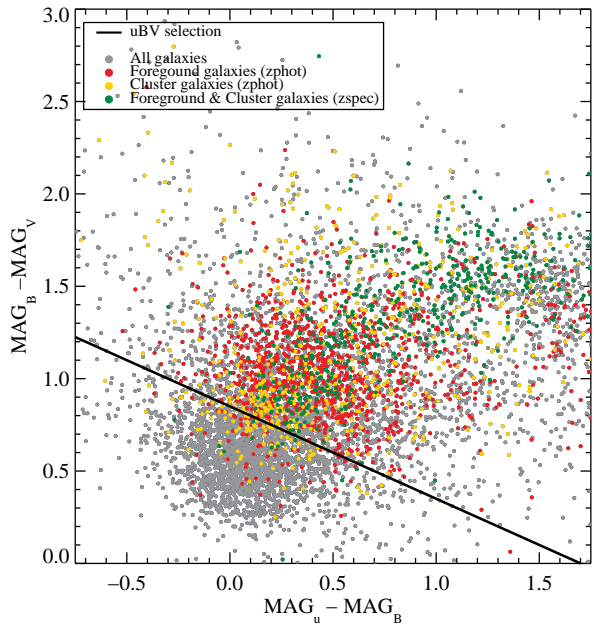}
\caption{As Fig.~\ref{fig:colorcolor0717} but for the B${-}$V and u${-}$B colours. The solid black line delineates the uBV colour-cut defined for this work to mitigate shear dilution by unlensed galaxies.}
\label{fig:colorcolor0717_UBV}
\end{figure}
%%%%%%%%%%%%%%%%%%%%%%%%%%%%%%%%%%%%

%______________________________________________________
% 3.2
%______________________________________________________
\subsection{Foreground, Cluster $\&$ Background Galaxy Identifications}
\label{sect:sec42}
Since only galaxies behind the cluster are gravitationally lensed, the presence of cluster members and foreground galaxies in our ACS catalogue dilutes the observed shear and reduces the significance of all quantities derived from it. Identifying and eliminating as many of the contaminating unlensed galaxies is thus critical.

As a first step, we identify cluster galaxies with the help of the catalogue of photometric and spectroscopic redshifts compiled by \cite{ma08} from groundbased observations of the MACSJ0717.5+3745 field; the limiting magnitude of this catalogue is $m_{R_{c}}=24$. According to \cite{ma10}, all galaxies with spectroscopic redshifts 0.522 $<$ $z_{spec}$ $<$ 0.566 and with photometric redshifts 0.48 $<$ $z_{phot}$ $<$ 0.61 can be considered to be cluster galaxies. An additional criterion can be defined using the photometric redshifts derived as described in Sect.~\ref{sect:sec23}. Taking into account the statistical uncertainty of  $\Delta$z = 0.021 of the photometric redshifts, galaxies are defined as cluster members if their photometric redshift satisfies the criterion:
$$|z_{phot} - z_{cluster}| < \sigma_{phot-z} ,$$
with
$$\sigma_{phot-z} = (1 + z_{cluster}) \Delta z = 0.036,$$
where $z_{phot}$ and $z_{cluster}$ are the photometric redshift of the galaxy and the spectroscopic redshift of the galaxy cluster respectively. Reflecting the need for a balance between completeness and contamination, these redshift limits are much more generous than those used in conjunction with spectroscopic redshifts, which set the redshift range for cluster membership to $3\sigma \sim 0.0122$. For more details, see C.-J. Ma (Ph.D.\ thesis), as well as \citet{ma08, ma10}.

%______________________________________________________
% Fig 5
%______________________________________________________
\begin{figure}%[!t]
\includegraphics[width=88mm]{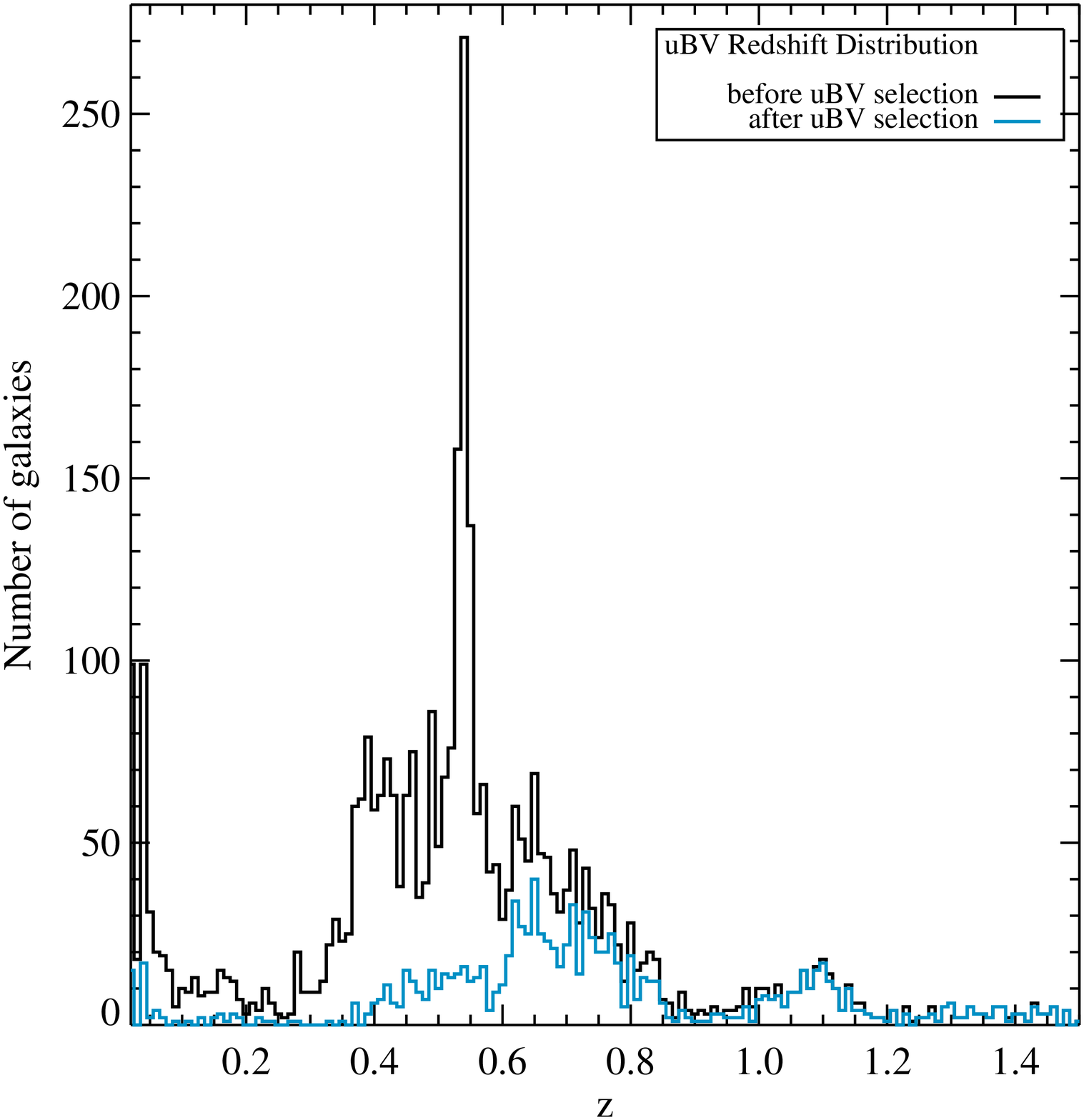}
\caption{Redshift distribution of all galaxies with u, B, and V photometry from CFHT/MegaCam and Subaru/SuprimeCam observations that have photometric or spectroscopic redshifts (black histogram). The cyan histogram shows the redshift distribution of galaxies classified as background objects using the uBV criterion illustrated in Fig.~\ref{fig:colorcolor0717_UBV}.}
\label{fig:Nzubv}
\end{figure}
%%%%%%%%%%%%%%%%%%%%%%%%%%%%%%%%%%%%

In spite of these cuts according to galaxy redshift, the remaining ACS galaxy sample is most likely still contaminated by foreground and cluster galaxies, the primary reason being the large difference in angular resolution and depth between the ACS and Subaru images. The relatively low resolution of the groundbased data causes the Subaru catalogue to be confusion limited and makes matching galaxies between the two catalogues difficult, especially near the cluster core. As a result, we can assign a redshift to only $\sim15$\% of the galaxies in the HST/ACS galaxy catalogue.

For galaxies without redshifts, we use colour-colour diagrams (B${-}$R vs R${-}$I, Fig.~\ref{fig:colorcolor0717}, and B${-}$V vs u${-}$B, Fig.~\ref{fig:colorcolor0717_UBV}) to identify foreground and cluster members. Using galaxies with spectroscopic or photometric redshifts from the full photometric Subaru catalogue with a magnitude limit of $m_{{\rm R}_{c}} = 25$,  we identify regions marked dominated by unlensed galaxies (foreground galaxies and cluster members). In the BRI plane we find ${\rm B}{-}{\rm R}<2.6 \,({\rm R}{-}{\rm I}) + 0.05$; $({\rm R}{-}{\rm I}) > 1.03$; or $({\rm B}{-}{\rm R}) < 0.9$ to best isolate unlensed galaxies; in the UBV plane the most efficient criterion is  ${\rm B}{-}{\rm V} < -0.5\, ({\rm u}{-}{\rm B}) + 0.85$.

Figures~\ref{fig:Nzbri} and \ref{fig:Nzubv} show the galaxy redshift distributions before and after the colour-colour cuts (BRI or uBV) are applied. The results of either kind of filtering are similar. The uBV selection is more efficient at removing cluster members and foreground galaxies at $z\le 0.6$ (20\% remain compared to 30\% for the BRI criterion) but also erroneously eliminates part of the background galaxy population. Comparing the convergence maps for both colour-colour selection schemes, we find the uBV selection to yield a better detection of structures in the area surrounding the cluster, indicating that suppressing contamination by unlensed galaxies is more important than a moderate loss of background galaxies from our final catalogue (see Sect.~\ref{sect:sec6} for more details).

Since the redshift distribution of the background population peaks at $0.61<z<0.70$ (cyan curve in Fig.~\ref{fig:Nzubv}) we assign, in the mass modelling phase, a redshift $z =0.65$ to background galaxies without redshift.

%______________________________________________________
% 3.3 :  
%______________________________________________________
\subsection{Shape measurements of Galaxies $\&$ Lensing Cuts}
%%%%%%%%%%%%%%%%%
\subsubsection{Theoretical Weak Gravitational Lensing Background}
\label{sect:sec331}
The shear signal contained in the shapes of lensed background galaxies is induced by a given foreground mass distribution. In the weak-lensing regime this shear is observed as a statistical deformation of background sources.
The observed shape of a source galaxy, $\varepsilon$, is directly related to the lensing-induced shear, $\gamma$, according to the relation :
$$\varepsilon = \varepsilon_{\rm intrinsic} + \varepsilon_{\rm lensing},$$
where $\varepsilon_{\rm intrinsic}$ is the intrinsic shape of the source galaxy (which would be observed in the absence of gravitational lensing), and 
$$\varepsilon_{\rm lensing} = \frac{\gamma}{1 - \kappa}.$$
Here $\kappa$ is the convergence. In the weak-lensing regime, $\kappa \ll 1$, which reduces the relation between the intrinsic and the observed shape of a source galaxy to
$$\varepsilon = \varepsilon_{\rm intrinsic} + \gamma.$$
Assuming galaxies are randomly oriented on the sky, the ellipticity of galaxies is an unbiased estimator of the shear, down to a limit referred to as ``intrinsic shape noise", $\sigma_{intrinsic}$  \citep[for more details see L07;][hereafter L10]{leauthaud10}. Unavoidable errors in the galaxy shape measurement are accounted for by adding them in quadrature to the``intrinsic shape noise":
$$\sigma_{\gamma}^{2} = \sigma_{\rm measurement}^{2} + \sigma_{\rm intrinsic}^{2}.$$

The shear signal induced on a background source by a given foreground mass distribution will depend on the configuration of the lens-source system. The convergence $\kappa$ is defined as the dimensionless surface mass density of the lens:
\begin{equation}
\kappa(\theta) =  \frac{1}{2} \nabla^{2}\varphi(\theta) = \frac{\Sigma(D_{OL}\theta)}{\Sigma_{\rm crit}},
\label{eqn:sigmacrit}
\end{equation}
where $\theta$ is the angular position of the background galaxy, $\varphi$ is the deflection potential, $\Sigma(D_{OL}\theta)$ is the physical surface mass density of the lens, and $\Sigma_{\rm crit}$ is the critical surface mass density defined as
$$\Sigma_{\rm crit} = \frac{c^{2} D_{OS}}{4 \pi G D_{OL} D_{LS}}.$$
Here, $D_{OL}$, $D_{OS}$, and $D_{LS}$ represent the angular distances from the observer to the lens, from the observer to the source, and from the lens to the source, respectively.

Considering the shear $\gamma$ as a complex number, we define
$$\gamma = \gamma_{1} + i \gamma_{2},$$
where $\gamma_{1} = |\gamma| \cos{2 \phi}$ and $\gamma_{2} = |\gamma| \sin{2\phi}$ are the two components of the shear, $\gamma$, defined previously, and $\phi$ is the orientation angle. With this definition, the shear is defined in terms of the derivatives of the deflection potential as :
$$\gamma_{1} = \frac{1}{2} (\varphi_{11} - \varphi_{22}),$$
$$\gamma_{2} = \varphi_{12} = \varphi_{21},$$
with
$$\varphi_{ij} = \frac{\partial^{2}}{\partial \theta_{i} \partial \theta_{j}} \varphi(\theta), \quad \quad i,j \in (1,2).$$

Following Kaiser $\&$ Squires (1993), the complex shear is related to the convergence by:
$$\kappa(\theta) = - \frac{1}{\pi} \int d^{2}\theta' \mathcal{R}e [\mathcal{D}(\theta - \theta') \gamma^{*}(\theta')].$$
Here $\mathcal{D}(\theta)$ is the complex kernel, defined as
$$\mathcal{D}(\theta) = \frac{\theta_{1}^{2} - \theta_{2}^{2} + 2i\theta_{1}\theta_{2}}{|\theta|^{4}},$$
and $\mathcal{R}e(x)$ defines the real part of the complex number $x$. The asterisk denotes complex conjugation.
The last equation shows that the surface mass density $\kappa(\theta)$ of the lens can be reconstructed straightforwardly if the shear $\gamma(\theta)$ caused by the deflector can be measured locally as a function of the angular position $\theta$.

%%%%%%%%%%%%%%%%%
\subsubsection{The RRG method}
\label{sect:sec332}
To measure the shape of galaxies we use the RRG method \citep{rhodes00} and the pipeline developed by L07. Having been developed for the analysis of data obtained from space, the RRG method is ideally suited for use with a small, diffraction-limited PSF as it decreases the noise on the shear estimators by correcting each moment of the PSF linearly, and only dividing them at the very end to compute an ellipticity.

The ACS PSF is not as stable as one might expect from a space-based camera. \cite{rhodes07} showed that both the size and the ellipticity pattern of the PSF varies considerably on time scales of weeks due to telescope 'breathing'. The thermal expansion and contraction of the telescope alter the distance between the primary and the secondary mirrors, inducing a deviation of the effective focus and thus from the nominal PSF which becomes larger and more elliptical. Using version 6.3 of the TinyTim ray-tracing program, \cite{rhodes07} created a grid of simulated PSF images at varying focus offsets.  By comparing the ellipticity of $\sim$ 20 stars in each image to these models,  \cite{rhodes07} were able to determine the effective focus of the images. Tests of this algorithm on ACS/WFC images of dense stellar fields confirmed that the best-fit effective focus can be repeatedly determined from a random sample of 10 stars brighter than $m_{\rm F814W} = 23$ with an rms error of $1\mu$m. Once images have been grouped by their effective focus position, the few stars in each images can be combined into one large catalog. PSF parameters are then interpolated using a polynomial fit in the usual weak-lensing fashion \citep{massey02}. More details on the PSF modelling scheme are given in \cite{rhodes07}.

The RRG method returns three parameters:  $d$, a measure of the galaxy size, and, the ellipticity represented by the vector $e = (e_{1},e_{2})$ defined as follows:
\begin{eqnarray*}
e &=& \frac{a^{2} - b^{2}}{a^{2} + b^{2}}\\ 
e_{1} &=& e \cos(2\phi) \\
e_{2} &=& e \sin(2\phi) ,
\end{eqnarray*}
where $a$ and $b$ are the half-major and half-minor axis of the background galaxy, respectively, and $\phi$ is the orientation angle of the ellipse defined previously. The ellipticity $e$ is then calibrated by a factor called shear polarizability, $G$, to obtain the shear estimator $\tilde{\gamma}$:
\begin{equation}
\tilde{\gamma} = C  \frac{e}{G}. \label{eqn:shearestimator}
\end{equation}
The shear susceptibility factor $G$ is measured from moments of the global distribution of $e$ and other shape parameters of higher order \citep[see ][]{rhodes00}. The Shear TEsting Program \citep[STEP; ][]{massey07} for COSMOS images showed that $G$ is not constant but varies as a function of redshift and S/N. To determine $G$ for our galaxy sample we use the same definition as the one used for the COSMOS weak-lensing catalogue (see L07):
$$G = 1.125 + 0.04 \arctan \frac{S/N - 17}{4}.$$
Finally, $C$, in Eq.~\ref{eqn:shearestimator} is the calibration factor. It was determined using a set of simulated images similar to those used by STEP \citep{heymans06, massey06} for COSMOS images, and is given by $C = (0.86_{-0.05}^{+0.07})^{-1}$ (for more details see L07).

%%%%%%%%%%%%%%%%%%%%%%
\subsubsection{Error of the Shear Estimator}
As explained in Sect.~\ref{sect:sec331}, the uncertainty in our shear estimator is a combination of intrinsic shape noise and shape measurement error:
$$\sigma_{\tilde{\gamma}}^{2} = \sigma_{\rm intrinsic}^{2} + \sigma_{\rm measurement}^{2},$$
where $\sigma_{\tilde{\gamma}}^{2}$ is referred to as shape noise. The shape measurement error is determined for each galaxy as a function of size and magnitude. Applying the method implemented in the PHOTO pipeline \citep{lupton01} to analyze data from the Sloan Digital Sky Survey, we assume that the optical moments of each object are the same as the moments computed for a best-fit Gaussian. Since the ellipticity components (which are uncorrelated) are derived from the moments, the variances of the ellipticity components can be obtained by linearly propagating the covariance matrix of the moments. The value of the intrinsic shape noise, $\sigma_{\rm intrinsic}$, is taken to be 0.27 (for more details see L07, L10).

In order to optimize the signal-to-noise ratio, we introduce an inverse-variance weighting scheme following L10:
$$w_{\tilde{\gamma}} = \frac{1}{\sigma_{\tilde{\gamma}}^{2}}.$$
Hence faint small galaxies which have large measurement errors are down-weighted with respect to sources that have well measured shapes.

%%%%%%%%%%%%%%%%%%%%%%
\subsubsection{Lensing Cuts}
The last step in constructing the weak-lensing catalogue for the MACSJ0717.5+3745 field consists of applying lensing cuts, i.e., to exclude galaxies whose shape parameters are ill-determined and will increase the noise in the shear measurement more than they add to the shear signal. However, in doing so, we need to take care not to introduce any biases. We use three galaxy properties to establish the following selection criteria: 
\begin{itemize}
\item[$\bullet$]
Their estimated detection significance:
$$\frac{S}{N} = \frac{FLUX\_AUTO}{FLUXERR\_AUTO} > 4.5;$$
where FLUX$\_$AUTO and FLUXERR$\_$AUTO are parameters returned by SE\textsc{xtractor};
\item[$\bullet$]
Their total ellipticity:
$$ e = \sqrt{e_{1}^{2} + e_{2}^{2}} < 1 ;$$
\item[$\bullet$]
Their size as defined by the RRG $d$ parameter:
$$d > 0.13\arcsec.$$
\end{itemize}

The requirement that the galaxy ellipticity be less than unity may appear trivial and superfluous. In practice it is meaningful though since the RRG method allows measured ellipticity values to be greater than 1 because of noise, although ellipticity is by definition restricted to $e\le 1$. Because Lenstool prevents ellipticities to be larger than 1, we removed the 251 objects with an ellipticity greater than unity from the RRG catalogue (2\% of the catalogue).
Serving a similar purpose, the restriction in the RRG size parameter $d$ aims to eliminate sources with uncertain shapes. PSF corrections become increasingly significant as the size of a galaxy approaches that of the PSF, making the intrinsic shape of a galaxy difficult to measure. 

Our final weak-lensing catalogue is composed of 10170 background galaxies, corresponding to a density of $\sim$ 52 galaxies arcmin$^{-2}$. In addition to applying the aforementioned cuts, and in order to ensure an unbiased mass reconstruction in the weak lensing regime only, we also remove all background galaxies located in the multiple-image (strong-lensing) region defined by an ellipse aligned with the cluster elongation and with a semi-major axis of 55", a semi-minor axis of 33". From the resulting mass-map presented in Sect.~\ref{sect:sec6}, we \emph{a-posteriori} derived a convergence histogram of the pixels outside this region, and found that 90\% of them are smaller than $\kappa = 0.1$, with a mean value of $\kappa = 0.03$. The shear and convergence are so weak because the ratio D$_{LS}$/D$_{OS}$ = 0.14 for background galaxies at redshift $z_{med} = 0.65$ and the cluster at $z = 0.54$ (see Sect.~\ref{sect:sec5}).

%______________________________________________________
% SECTION 4 :  DISTRIBUTION OF MASS
%______________________________________________________
\section{Mass Distribution}
\label{sect:sec5}
The mass modelling for the entire MACSJ0717.5+3745 field is performed using the LENSTOOL\footnote{LENSTOOL is available online: http://lamwws.oamp.fr/lenstool} \citep{jullo07} software, using the adaptive-grid technique developed by \cite{jullo09} and modified by us for weak-lensing mass measurements. Because LENSTOOL implements a Bayesian sampler, it provides many mass maps fitting the data that  can be used to obtain a mean mass map and to determine its error.

%______________________________________________________
% 4.1
%______________________________________________________
\subsection{Multi-Scale Grid Method}
\label{sect:sec51}
We start with the method proposed by \cite{jullo09} to model the  cluster mass distribution using gravitational lensing. This recipe uses a multi-scale grid of Radial Basis Functions (RBFs) with physically motivated profiles and lensing properties. With a minimal number of parameters, the grid of RBFs of different sizes provides higher resolution and sharper contrast in regions of higher density where the number of constraints is generally higher.
It is well suited to describe irregular mass distributions like the one investigated here.

The initial multi-scale grid is created from a smoothed map of the cluster K-band light and is recursively refined in the densest regions. In the case of MACSJ0717.5+3745, this method is fully adaptive as we want to sample a wide range of masses, from the cluster core to the far edge of the HST/ACS field where the filamentary structure is least dense. Initially, the field of interest is limited to a hexagon, centred on the cluster core and split into six equilateral triangles \citep[see Fig. 1 in][]{jullo09}. This initial grid is subsequently refined by applying a splitting criterion that is based on the surface density of the light map. Hence, a triangle will be split into four smaller triangles if it contains a single pixel that exceeds a predefined light-surface-density threshold.

Once the adaptive grid is set up, RBFs described by Truncated Isothermal Mass Distributions (TIMD), circular versions of truncated Pseudo Isothermal Elliptical Mass Distributions (PIEMD) \citep[see, e.g.,][]{kassiola93,kneib96,limousin05,eliasdottir07} are placed at the grid node locations.
The analytical expression of the TIMD mass surface density is given by
$$
\Sigma(R) = \sigma_{0}^{2} f(R, r_{\rm core}, r_{\rm cut})
$$
with
$$
f(R, r_{\rm core}, r_{\rm cut}) = \frac{1}{2G} \frac{r_{\rm cut}}{r_{\rm cut} - r_{\rm core}} \left(\frac{1}{\sqrt{r_{\rm core}^{2} + R^{2}}} - \frac{1}{\sqrt{r_{\rm cut}^{2} + R^{2}}}\right) .
$$

Hence, $f$ defines the profile, and $\sigma_{0}^{2}$ defines the weight of the RBF. This profile is characterized by two changes in slope at radius values of $r_{\rm core}$ and $r_{\rm cut}$. Within $r_{\rm core}$, the surface density is approximately constant, between $r_{\rm core}$ and $r_{\rm cut}$, it is isothermal (i.e., $\Sigma \propto r^{-1}$), and beyond $r_{\rm cut}$ it falls as $\Sigma \propto r^{-3}$.
This profile is physically motivated and meets the three important criteria of a) featuring a finite total mass, b) featuring a finite central density,  and c) being capable of describing extended flat regions, in particular in the centre of clusters.

The RBFs' $r_{\rm core}$ value is set to the size of the smallest nearby triangle, and their $r_{\rm cut}$ parameter is set  to $3r_{\rm core}$, a scaling that \cite{jullo09} found to yield an optimal compromise between model flexibility and overfitting. We then adapt the technique proposed by \cite{diego07} to our multiscale grid model to fit the weak-lensing data ($\kappa \ll 1$). Assuming a set of $M$ images and a model comprised of $N$ RBFs, the relation between the weights ${\sigma_{0}^{2}}^{i}$ and the 2$M$ components of the shear is given by

$$
\left[\begin{array}{c}  
\gamma_{1}^{1} \\
\vdots \\
\gamma_{1}^{M} \\
\gamma_{2}^{1} \\
\vdots \\
\gamma_{2}^{M}
\end{array}\right] 
= 
\left[\begin{array}{ccc}
\Delta_{1}^{(1,1)} & \cdots & \Delta_{1}^{(1,N)} \\
\vdots & \ddots & \vdots \\\nonumber
\Delta_{1}^{(M,1)} & \cdots & \Delta_{1}^{(M,N)} \\
\Delta_{2}^{(1,1)} & \cdots & \Delta_{2}^{(1,N)} \\
\vdots & \ddots & \vdots \\
\Delta_{2}^{(1,M)} & \cdots & \Delta_{2}^{(M,N)}
\end{array}\right]
\left[\begin{array}{c}
{\sigma_{0}^{2}}^{1}\\
\vdots \\
{\sigma_{0}^{2}}^{N} 
\end{array}\right] 
$$
with 
$$
\Delta_{1}^{(i,j)} = \frac{D_{LS}^{i}}{D_{OS}^{i}} \gamma_{1}^{(i,j)},
$$
$$
\Delta_{2}^{(i,j)} = \frac{D_{LS}^{i}}{D_{OS}^{i}} \gamma_{2}^{(i,j)},
$$
where 
$$
\gamma_{1}^{(i,j)} = \frac{1}{2} (\partial_{11}\Phi_{j}(R_{ij}) - \partial_{22}\Phi_{i}(R_{ij})) ,
$$
$$
\gamma_{2}^{(i,j)} = \partial_{12}\Phi_{j}(R_{ij}) = \partial_{21}\Phi_{j}(R_{ij}) ,
$$
and 
$$
R_{ij} = | \theta_{i} - \theta_{j} |.
$$

%______________________________________________________
% Fig 6
%______________________________________________________
\begin{figure}
\includegraphics[width=84mm]{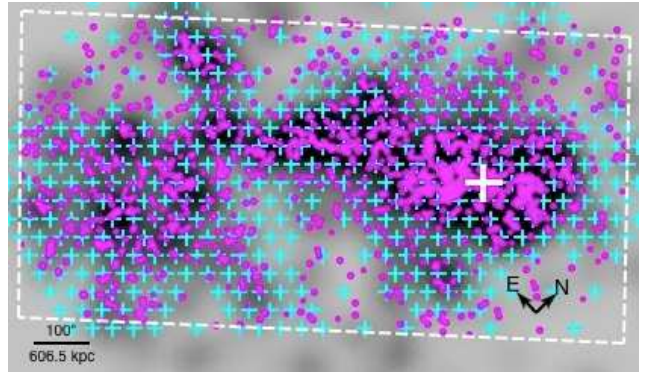}
\caption{Distribution of grid nodes and galaxy-scale potentials, superimposed on the K-band light map of MACSJ0717.5+3745. Cyan crosses represent the location of 468 RBFs at the nodes of the multi-scale grid; magenta circles represent galaxy-scale potentials used to describe the contribution of individual cluster members. The white dashed line defines the HST/ACS field of view. The white cross marks the cluster centre at 07:17:30.025, +37:45:18.58 (R.A., Dec).} 
\label{fig:potgrid}
\end{figure}
%%%%%%%%%%%%%%%%%%%%%%%%%%%%%%%%%%%%

In the above, $D_{LS}^{i}$ and $D_{OS}^{i}$ are the angular distances from the RBF $j$ to the background source $i$ and from the observer to the background source $i$, respectively, and $\Phi$ is the projected gravitational potential. 
$\gamma_{k=1,2}$ are the two components of the shear, $\Delta_{k=1,2}^{(i,j)}$ is the value of the RBF normalized with ${\sigma_{0}^{2}}^{j} = 1$, centered on the grid node located at $\theta_{j}$, and computed at a radius $R = |\theta_{i} - \theta_{j}|$. The contribution of this RBF to the predicted shear at location $\theta_{i}$ is given by the product $\Delta_{k=1,2}^{(i,j)} {\sigma_{0}^{2}}^{j}$ (see Eq.~\ref{eqn:sigmacrit}).
More details about this method will be provided in a forthcoming paper  (Jullo et al. 2012, in preparation).

Figure~\ref{fig:potgrid} shows the distribution of the 468 RBFs superimposed on the light map used to define the multi-scale grid. The smallest RBF has a $r_{core} = 26\arcsec$. Note that although Lenstool can perform strong lensing, and reduced-shear optimization, here the proposed formalism assumes weak-shear with $\kappa \ll 1$. To get an unbiased estimator, we remove the background galaxies near the cluster core from the catalogue, but we keep RBFs in this region as we found it improves the mass reconstruction. In order to check our weak-shear assumption, we multiply the resulting convergence map by the factor $1 - \kappa$, and find this minor correction only represents about 5$\%$ at 300~kpc and less than 1$\%$ at 500~kpc. It confirms that working with a weak-shear assumption in this case is not biasing our results. On Fig.~\ref{fig:massprof_slwl}, the black dashed curve corresponds to the corrected  weak-shear optimization, while the black curve represents the weak-shear optimization profile. Both shows a really good agreement with the strong-lensing results (magenta curve).

%______________________________________________________
% 4.2
%______________________________________________________
\subsection{Cluster Member Galaxies}
Our catalogue of cluster members is compiled from the photometric catalogue of \cite{ma08}. Within the HST/ACS study area we identify  as cluster members 1244 galaxies with spectroscopic and/or photometric redshifts within the redshift ranges defined in Sect.~\ref{sect:sec42}. We use these galaxies' $K_{S}$-band luminosities as mass estimators (see below for details).

All cluster members are included in the lens model in the form of truncated PIEMD potentials (see Sect.~\ref{sect:sec51}) with characteristic properties scaled according to their K-band luminosity:
$$r_{\rm core} = r^{*}_{\rm core} \left(\frac{L}{L^{*}}\right)^{1/2},$$
$$r_{\rm cut} = r^{*}_{\rm cut} \left(\frac{L}{L^{*}}\right)^{1/2},$$
and,
$$\sigma_{0} = \sigma^{*}_{0} \left(\frac{L}{L^{*}}\right)^{1/4} .$$

These scaling relations are found to well describe early-type cluster galaxies \citep[e.g.][]{wuyts04,fritz05} under the  assumptions of mass tracing light and the validity of the \cite{FJ76} relation. Since the mass, $M$, is proportional to $\sigma^{2}_{0} r_{\rm cut}$, the above relations ensure $M \propto L$, assuming that the mass-to-light ratio is constant for all cluster members. 

To find suitable values of $r^{*}_{\rm core}$, $r^{*}_{\rm cut}$, and $\sigma^{*}_{0}$ we take advantage of the results of the strong-lensing analysis of MACSJ0717.5+3745 recently conducted by \cite{limousin11}. Their mass model of the cluster also included cluster members, using the scaling relations defined above \citep[see also][]{limousin07b}. For a given $L^{*}$ luminosity, given by $m^{*} = 19.16$, the mean apparent magnitude of a cluster member in K-band, \cite{limousin11} set all geometrical galaxy parameters (centre, ellipticity, position angle) to the values measured with SE\textsc{xtractor}, fixed $r^{*}_{core}$ at 0.3~kpc, and then searched for the values of $\sigma^{*}_{0}$ and $r^{*}_{\rm cut}$  that yield the best fit. We here use the same  best-fit values, $r_{cut}^{*} = 60$~kpc and $\sigma_{0}^{*} = 163$~kpc/s, to define the potentials of the cluster members. Hence, all parameters describing cluster members are fixed in our model; their positions are marked by the magenta circles in Fig.~\ref{fig:potgrid}.

%______________________________________________________
% Fig 7
%______________________________________________________
\begin{figure}
\includegraphics[width=84mm]{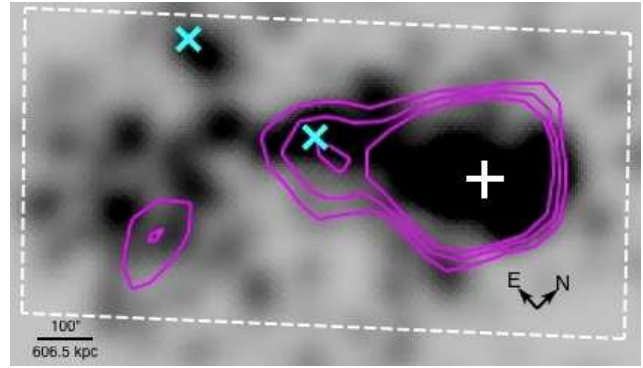}
\caption{Contours of the convergence $\kappa$ in the MACS0717.5+3745 field obtained using the inversion method described by Seitz $\&$ Schneider (1995), overlaid on the K-band light map. Magenta contours represent the 1, 2 and 3$\sigma$ contours. Cyan crosses mark the position of  two galaxy groups (see text for details). The white cross marks again the cluster centre (cf.\ Fig.~\ref{fig:potgrid}).} 
\label{fig:kappamap}
\end{figure}
%%%%%%%%%%%%%%%%%%%%%%%%%%%%%%%%%%%%

%______________________________________________________
% 4.3
%______________________________________________________
\subsection{Mass Modeling}
The mass reconstruction is conducted using LENSTOOL which implements an optimisation method based on a Bayesian Markov Chain Monte Carlo (MCMC) approach \citep{jullo07}. We chose this approach because we want to propagate as transparently as possible errors on the ellipticity into errors on the filament mass measurement.
This method provides two levels of inference: parameter space exploration and model comparison, by means of the posterior Probability Density Function (PDF) and the Bayesian evidence, respectively.

All of these quantities are related by the Bayes Theorem:
$$ {\rm Pr}(\theta|D,M) \propto \frac{{\rm Pr}(D|\theta,M) {\rm Pr}(\theta|M)}{Pr(D|M)}, $$
where Pr($\theta|D,M$) is the posterior PDF, Pr($D|\theta,M$) is the likelihood of a realisation yielding the observed data $D$ given the parameters $\theta$ of the model $M$, and, Pr($\theta|M$) is the prior PDF for the parameters.
Pr($D|M$) is the probability of obtaining the observed data $D$ given the assumed model $M$, also called the Bayesian evidence. It measures the complexity of the model. The posterior PDF will be the highest for the set of parameters $\theta$ that yields the best fit and is consistent with the prior PDF. \cite{jullo07} implemented an annealed Markov Chain to converge progressively from the prior PDF to the posterior PDF.

For the weak-lensing mass mapping, we implement an additional level of  complexity based on the Gibbs sampling technique \citep[see Massive Inference in the Bayesys manual,][]{skilling98}. Basically, only the posterior distribution of the most relevant RBFs is explored. The number of RBFs to explore is an additional free parameter with a Poisson prior. Exploring possible values of this prior in simulations, we find that the input mass is well recovered when  the mean of this prior is set to 2$\%$ of the total number of RBFs in the model. 
The weights of the RBFs, ${\sigma_{0}^{2}}^{i}$, are decomposed into the product of a quantum element of weight, $q$, common to all RBF, and a multiplicative factor $\zeta^{i}$. In order to have positive masses, we make $\zeta$ follow a Poisson prior (case MassInf=1), and $q$ to follow the prior distribution $\pi(q) = q_{0}^{-2} q \exp^{-q/q_{0}}$, and we fix the initial guess $q_{0} = 10$~km$^{2}$/s$^{2}$.
The final distribution of ${\sigma_{0}^2}^i$ is well approximated by a distribution $\pi(q)$ with $q_{0} = 12$~km$^{2}$/s$^{2}$, and, we find that $16.8\pm4.8$ RBFs are necessary on average to reconstruct the mass distribution given our data, i.e. about 3.5$\%$ of the total number of RBFs in the model. This new algorithm is fast, as it can deliver a mass map in 20 minutes for about 10,000 galaxies and 468 RBFs --- the previous algorithm used in \citet{jullo09} was taking more than 4 weeks to converge. The new algorithm has been tested in parallel on five processors clocked at 2 GHz. 

We define the likelihood function as (see e.g. \cite{schneider00}:
$${\rm Pr}(D|\theta) = \frac{1}{Z_{L}} \exp{(- \frac{ \chi^2}{2})},$$
where $D$ is the vector of ellipticity component values, and $\theta$ is a vector of the free parameters $\sigma_{i}^{2}$. $\chi^2$ is the usual goodness-of-fit statistic:
\begin{equation}
\chi^2 = \sum_{i=1}^{M} \sum_{j=1}^{2} \frac{(\tilde{\gamma}_{j,i} - 2 \gamma_{j,i}(\theta_{i}))^2}{\sigma_{\tilde{\gamma}}^2}. \label{eqn:chi2}
\end{equation}
The $2M$ intrinsic ellipticity components ($M$ is the number of background sources), are defined as follows:
\begin{equation}
\varepsilon_{{\rm intrinsic}, j} = \tilde{\gamma}_{j}(\theta_{j}) - 2 \gamma_{j}(\theta_{j}) \label{eqn:intrinsic}
\end{equation}
and, are assumed to have been drawn from a Gaussian distribution with variance defined in Sect.~\ref{sect:sec331} :
$$\sigma_{\tilde{\gamma}}^{2} = \sigma_{\it intrinsic}^{2} + \sigma_{measure}^{2}.$$
The normalization factor is given by
$$ Z_{L} = \prod_{i=1}^{M} \sqrt{2 \pi} \,\, \sigma_{\tilde{\gamma}_{i}}.$$

%______________________________________________________
% Fig 8
 %______________________________________________________ 
\begin{figure}
\includegraphics[width=84mm]{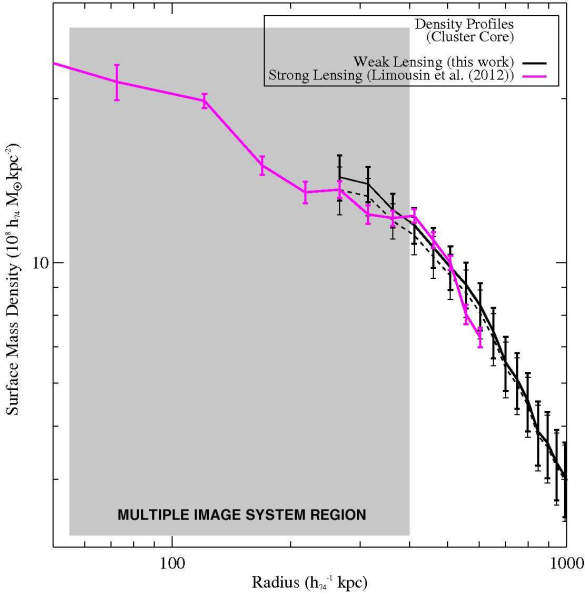}
\caption{Density profiles within the core of MACSJ0717.5+3749. The magenta curve represents strong-lensing results from \citet{limousin11}; the black curve shows the profile derived from our weak-lensing  analysis; the black dashed curve shows the weak-lensing profile corrected from the weak-shear approximation}. The grey area marks the region within which multiple images are observed.
\label{fig:massprof_slwl}
\end{figure}
%%%%%%%%%%%%%%%%%%%%%%%%%%%%%%%%%%%%%%

Note the factor of 2 in Eqn.~\ref{eqn:chi2} and~\ref{eqn:intrinsic}  because of the particular definition of the ellipticity in RRG (see Sect.~\ref{sect:sec332}). The logarithmic Bayesian evidence is then given by
$$ \log(E) = - \frac{1}{2} \int_0^1 \langle \chi^2 \rangle ^\lambda d\lambda$$
where the average is computed over a set of 10 MCMC realizations at any given iteration step $\lambda_i$, and the integration is performed over all iterations $\lambda_i$ from the initial model ($\lambda=0$) to the best-fit result ($\lambda=1$). An increment in $d\lambda$ depends on the variance between the 10 likelihoods computed at a given iteration, and a convergence rate that we set equal to 0.1 (see Bayesys manual for details). The increment gets larger as the algorithm converges towards 1.

%______________________________________________________
% Section 5 :  Results
%______________________________________________________
\section{Results}
\label{sect:sec6}

%______________________________________________________
% Fig 9
%______________________________________________________
\begin{figure}
\includegraphics[width=84mm]{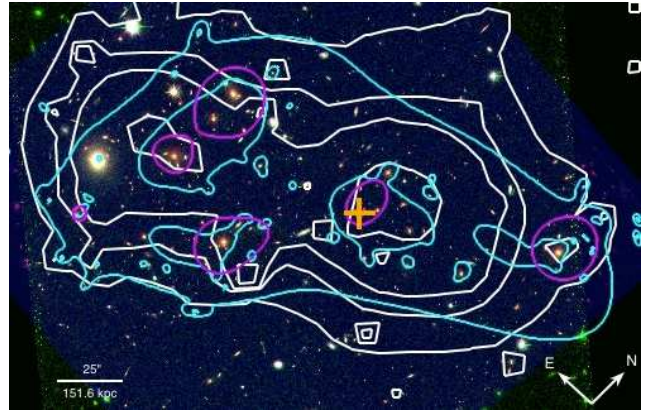}
\caption{Mass distribution in the core of MACSJ0717.5+3745. Cyan contours represent the mass distribution inferred in the strong-lensing study by \citet{limousin11}; white contours show the mass distribution obtained by our weak-lensing analysis; and magenta contours represent the distribution of the cluster K-band light. The orange cross marks the cluster center adopted in the analysis (cf.\ Fig.~\ref{fig:potgrid}).}
\label{fig:contours_slwl}
\end{figure}
%%%%%%%%%%%%%%%%%%%%%%%%%%%%%%%%%%%%%%

%______________________________________________________
% 5.1
%______________________________________________________
\subsection{Kappa Map : Standard mass reconstruction} \label{sect:kappamap}
We test our catalogue of background galaxies by mapping the convergence $\kappa$ in the MACSJ0717.5+3745 field. We use the method described in Sect.~\ref{sect:sec331}, which is based on the inversion equation found by Kaiser $\&$ Squires (1993) and developed further by \cite{seitz95}. It shows that the best density reconstructions are obtained when the smoothing scale is adapted to the strength of the signal.  We find  a smoothing scale of 3~arcmin to provide a good compromise between signal-to-noise considerations and map resolution. We estimate the noise directly from the measured errors, $\sigma_{\rm measurement}$, averaged within the grid cells.

The resulting $\kappa$-map is shown in Fig.~\ref{fig:kappamap}, overlaid on a smoothed image of the K-band light from cluster galaxies. We identify two substructures south-east of the cluster core whose locations match the extent of the filamentary structure seen in the galaxy distribution. The first of these, near the cluster core, coincides with the beginning of the optical filament; the second falls close to the apparent end of the filament close to the edge of our study area. A simple inversion technique thus already yields a 2~sigma detections of parts of the filament.

%______________________________________________________
% Fig 10
%______________________________________________________
\begin{figure*}
\includegraphics[width=178mm]{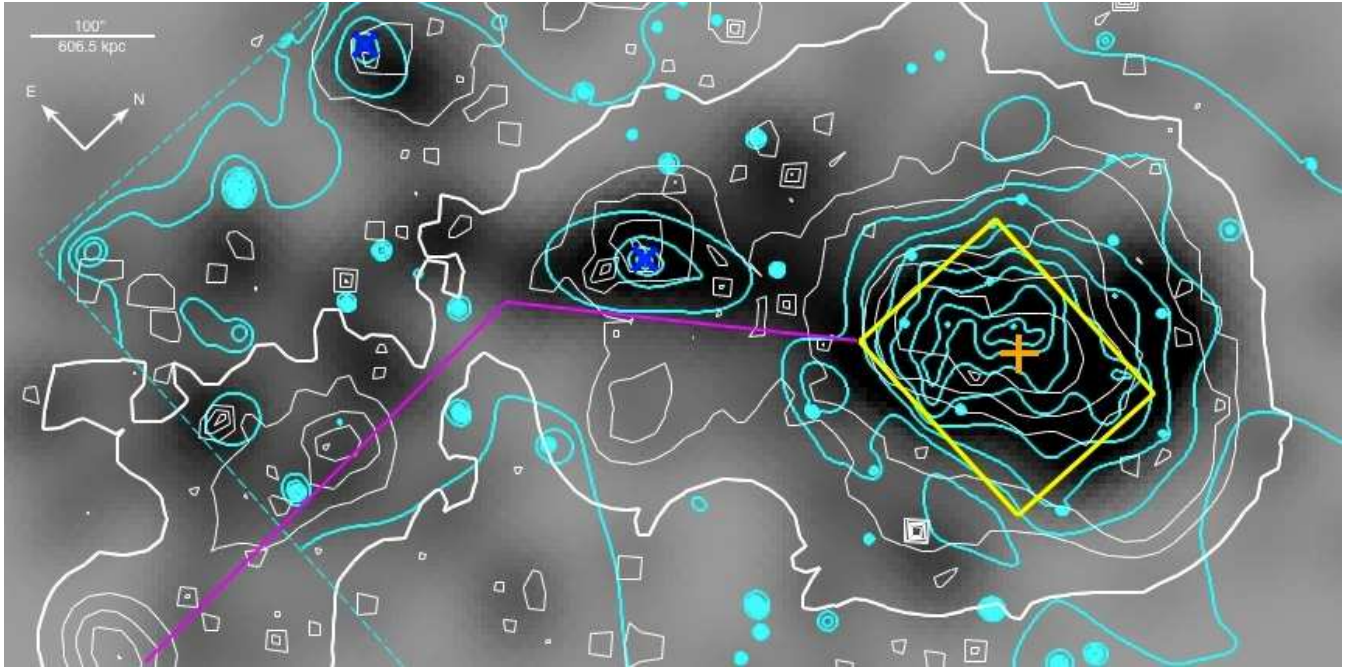}
\caption{
Mass map from our weak-gravitational lensing analysis overlaid on the light map of the mosaic.
The two sets of contours show the X-ray surface brightness (cyan), and weak-lensing mass (white). The bold white contour corresponds to a density of $1.84\times 10^{8}\,h_{74}$ M$_{\odot}$.kpc$^{-2}$. The orange cross marks the location of the fiducial cluster centre, and the two blue crosses show the positions of the two X-ray detected satellite groups mentioned in Sect.~\ref{sect:kappamap}. The dashed cyan lines delineates the edges of the Chandra/ACIS-I field of view. The yellow box, finally, marks the cluster core region shown in Fig.~\ref{fig:contours_slwl}. The magenta line emphasises the extent and direction of the large-scale filament.}
\label{fig:massmap_WL}
\end{figure*}
%%%%%%%%%%%%%%%%%%%%%%%%%%%%%%%%%%%%%%

The Chandra observation of MACS0717.5+3745 shows X-ray detections of two satellite groups of galaxies embedded in the filament \citep[private communication H. Ebeling, see also Fig.~1 of][and Fig.~\ref{fig:massmap_WL} of this paper]{ma09}. Their X-ray coordinates (R.A., Dec, J2000) are : i) 07:17:53.618, +37:42:10.46, and, ii) 07:18:19.074, +37:41:13.14 (cyan crosses in Fig.~\ref{fig:kappamap}).
The first of these, close to the filament-cluster interface, is detected by the inversion technique, but not the second. Conversely, the peak  in the convergence map in the southeastern corner of our study area is not detected in X-rays.

%______________________________________________________
% 5.2
%______________________________________________________
\subsection{Mass Map : Iterative mass reconstruction}
The primary goal of this paper is the detection of the large-scale filament with weak-lensing techniques and the characterisation of its mass content. To calibrate the mass map of the entire structure (i.e., filament + cluster core) and overcome the mass-sheet degeneracy, we compare the weak-lensing mass obtained for the cluster core with  the technique described in Sect.~\ref{sect:sec5}, with the strong lensing (SL) results from \cite{limousin11}. 

%%%%%%%%%%%%%%%%%%%%%%%%%%%%%%%%%
\subsubsection{Cluster Core}
\cite{limousin11} inferred a parametric mass model for the core of MACSJ0717.5+3745 using strong-lensing (SL) constraints, specifically 15 multiple-image systems identified from multi-color data within a single HST/ACS tile. Spectroscopic follow up of the lensed features allowed the determination of a well calibrated mass model.
The cyan contours in Fig.~\ref{fig:contours_slwl} show the mass distribution obtained from their analysis.
Four mass components, associated with the main light components, are needed to satisfy the observational constraints.  The cluster mass  reported by  \cite{limousin11} for the region covered by a single ACS field-of-view is $M_{\rm SL}(R<500\, {\rm kpc}) = (1.06 \pm 0.03)~10^{15}$~$h_{74}^{-1}$ M$_{\odot}$ where the cluster centre is taken to be at the position quoted before (07:17:30.025  +37:45:18.58). This position was adopted as the cluster centre because it marks the barycenter of the Einstein ring measured by \cite{meneghetti11}. Another SL analysis of MACSJ0717.5+3745 was performed by \cite{zitrin09b} who report a mass of $M_{\rm SL} (R<350\,{ kpc}) = (7.0 \pm 0.5)~10^{14}$~$h_{74}^{-1}$ M$_{\odot}$.

Figure~\ref{fig:massprof_slwl} compares the mass density profiles derived from the SL analysis (magenta line) and from our WL analysis (black line); note the very good agreement. The WL density profile has been cut  up to $\sim$300~kpc as this region corresponds to the multiple-image regime, therefore does not contain any WL constraints.

In Fig.~\ref{fig:massprof_slwl}, the SL density profile is extrapolated beyond the multiple-image region (the cluster core), while the WL density profile is extrapolated into the multiple-image region. The WL mass thus obtained for the cluster core is $M_{\rm WL}(R<500\, {\rm kpc}) = (1.04 \pm 0.08)~10^{15}$ $h_{74}^{-1}$ M$_{\odot}$ in excellent agreement with the value measured by \cite{limousin11}. We also measure $M_{\rm WL}(R<350\, {\rm kpc}) = (5.10 \pm 0.54)~10^{14}$ $h_{74}^{-1}$ M$_{\odot}$, in slight disagreement with the result reported by \cite{zitrin09b}.
But the excellent agreement at larger radii between strong- and weak-lensing results without any adjustments validates our redshift distribution for the background sources, and our new reconstruction method.

Figure~\ref{fig:contours_slwl}  compares the mass contours from the two different gravitational-lensing analyses for the core of MACSJ0717.5+3745 and illustrates the remarkable agreement between these totally independent measurements. Our WL analysis requires a bi-modal mass concentration which coincides with the two main concentrations inferred by the SL analysis of \cite{limousin11}. 

%%%%%%%%%%%%%%%%%%%%%%%%%%%%%%%%%%
\subsubsection{Detection of a filamentary structure}
Figure~\ref{fig:massmap_WL} shows the mass map obtained for the whole HST/ACS mosaic using the modeling and  optimization method described in Sect.~\ref{sect:sec5}. Also shown are  the X-ray surface brightness as observed with Chandra/ACIS-I, and the cluster K-band light. A filamentary structure extending from the south-eastern edge of the cluster core to the south-eastern edge of the ACS mosaic area is clearly visible in all three datasets. The projected length of the filament is measured as $4.5\,h_{74}^{-1}$ Mpc, and the mean density in the region of the filament is found to be  $(2.92 \pm 0.66) \times 10^{8}\,h_{74}$ M$_{\odot}$ kpc$^{-2}$. The mass concentration detected to the South-East of the cluster core corresponds to a low-mass structure comparable to a galaxy group and indeed coincides with a poor satellite cluster of MACSJ0717.5+3745 that has been independently identified as an overdensity of galaxies and as a diffuse peak in the X-ray surface brightness. The second, more compact, X-ray emitting galaxy group (almost due East of the cluster core and outside the filament) is also detected. 

To assess the validity of this filament detection, we test the robustness of our multi-scale grid optimization method.
As a first step we create a uniform grid with a RBF of size $r_{core} = 21\arcsec$ and 3169 potentials.  We found that using this uniform grid added noise in locations where no structures were detected neither in the optical, nor in the mass map produced with the multi-scale grid. The Bayesian evidence obtained for the uniform grid (pixel size of $r_{core} = 21\arcsec$) is $\log(E) = -6589$, compared to $\log(E) = -6586$ for the multi-scale grid (pixel size of $r_{core} = 26\arcsec$). The small difference between these two values demonstrates that the reduced number of RBFs in our multi-scale grid does not compromise the measurement. This is further confirmed by a test using a low-resolution uniform grid with a pixel size of $r_{core} = 43\arcsec$ comprising 817 RBFs, for which we found a Bayesian evidence of $\log(E) = -6582$. All 3 grids equivalently reproduce the data. Our multi-scale grid model provides an intermediate resolution map.

Having established that the use of an adaptive grid has no detrimental effect on the Bayesian evidence, the second test consisted in increasing the resolution of the multi-scale grid to a pixel size of $r_{core} = 13\arcsec$, resulting in 2058 potentials. We found that the noise increased with the resolution, whereas all real mass concentrations maintained their size. The Bayesian evidence obtained with this high-resolution multi-scale grid is $\log(E) = -6608$. A WL mass reconstruction with the high-resolution multi-scale grid recovers the density profile obtained with the multi-scale grid using $r_{core} = 26\arcsec$ pixels. 
However, in view of the large difference between these two evidence values, it is clear that such an increase in resolution is not required by the data.

%______________________________________________________ 
% SECTION 6 :  PROPERTIES OF THE FILAMENT
%______________________________________________________
\section{Properties of the FIlament}
\label{sect:sec7}
In this Section, we first summarise briefly the results of previous studies of the MACSJ0717.5+3745 filament, and then discuss the filament's properties as derived from our weak-lensing analysis.
 
The detection of a filamentary structure in the field of the cluster MACSJ0717.5+3745 was first reported by \cite{ebeling04}, who discovered a pronounced overdensity of galaxies with  V$-$R colours close to the cluster red sequence, extending over  $\sim 4 h_{74}^{-1}$ Mpc at the cluster redshift. Extensive spectroscopic follow-up of over 300 of these galaxies in a region covering both the cluster and the filament confirmed that the entire structure is located at the cluster redshift, $z = 0.545$. This direct detection of an extended large-scale filament connected to a massive, distant cluster lent strong support to predictions from theoretical models and numerical simulations of structure formation in a hierarchical scenario, according to which large-scale filaments funnel matter toward galaxy clusters inhabiting the nodes of the cosmic web. Taking advantage of the opportunity provided by this special target,
\cite{ma08} pursued a wide-field spectroscopic analysis of the galaxy population of both the cluster and the filament. Along the filament, they found a significant offset in the average redshift of galaxies, corresponding to $\sim 630$ km s$^{-1}$ in velocity, as well as a decrease in velocity dispersion from the core of the cluster to the end of the filament. 
\cite{ma10} studied the morphology of cluster members across a $100\, h_{74}^{-2}$~Mpc$^{2}$ area to investigate the effect of cluster environment on star formation using HST/ACS data.  They explored the relation between galaxy color and density. They showed that the 1-D density profile of MACSJ0717.5+3745 cluster members followed the filament from the core to beyond the virial radius falling, then slightly rising and flattening at $\sim 2 h_{74}^{-1}$~Mpc.

%______________________________________________________
% Fig 11
%______________________________________________________
 \begin{figure}
\includegraphics[width=84mm]{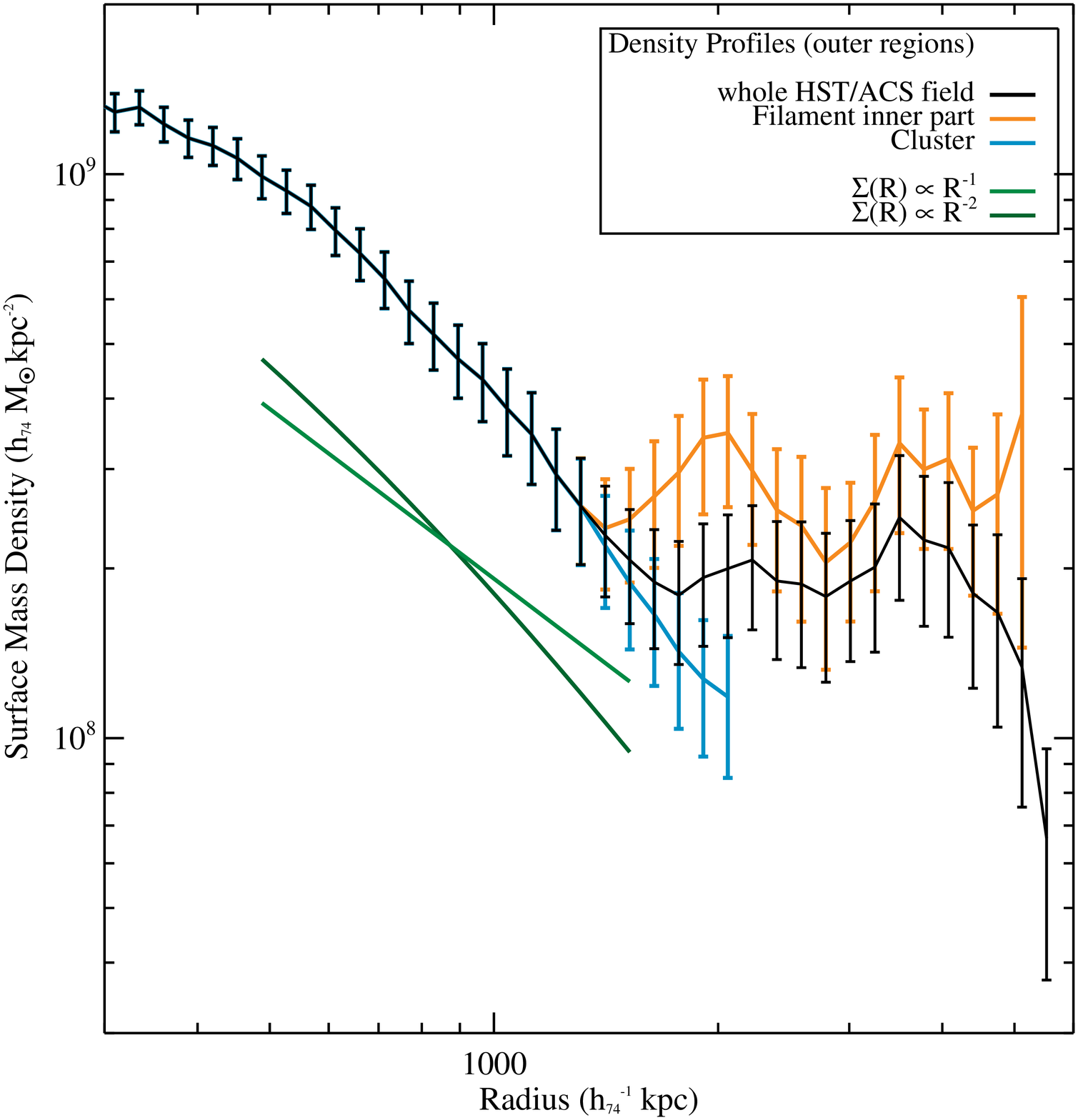}
\caption{Profile of the mass surface density of the cluster/filament complex as a function of distance from the cluster core. The three curves show the mass surface density measured in our weak-lensing analysis across the entire field (black), for the cluster without the filament (cyan), and for the inner part of the filament (orange). The two green lines represent two different slopes: $R^{-1}$ and $R^{-2}$. See text for details.}
\label{fig:densprof_fit}
\end{figure}
%%%%%%%%%%%%%%%%%%%%%%%%%%%%%%%%%%%%%%

\begin{figure}
\includegraphics[width=84mm]{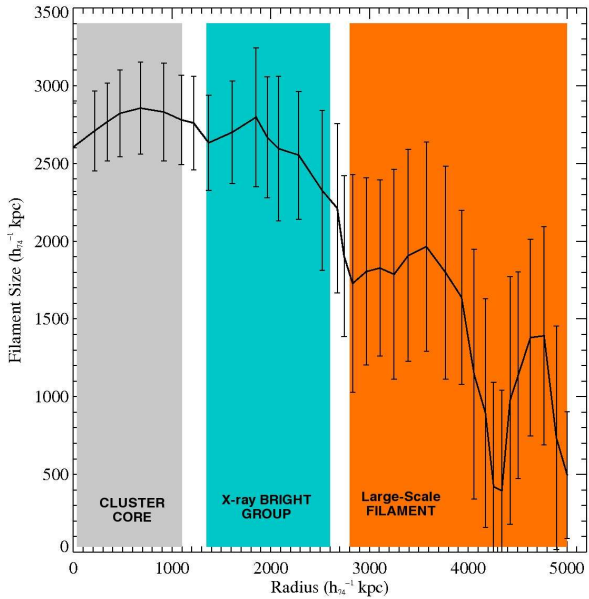}
\caption{Diameter of the filamentary structure as a function of distance from the cluster centre. The regime of the cluster proper and the segment containing the embedded X-ray bright group of galaxies (blue cross in Fig.~\ref{fig:massmap_WL}) are marked. The density threshold used to define the filamentary structure is $1.84\times 10^{8}\,h_{74}$ M$_{\odot}$.kpc$^{-2}$.} 
\label{fig:r_filament}
\end{figure}

\begin{figure}
\hspace{-2mm}\includegraphics[width=90mm]{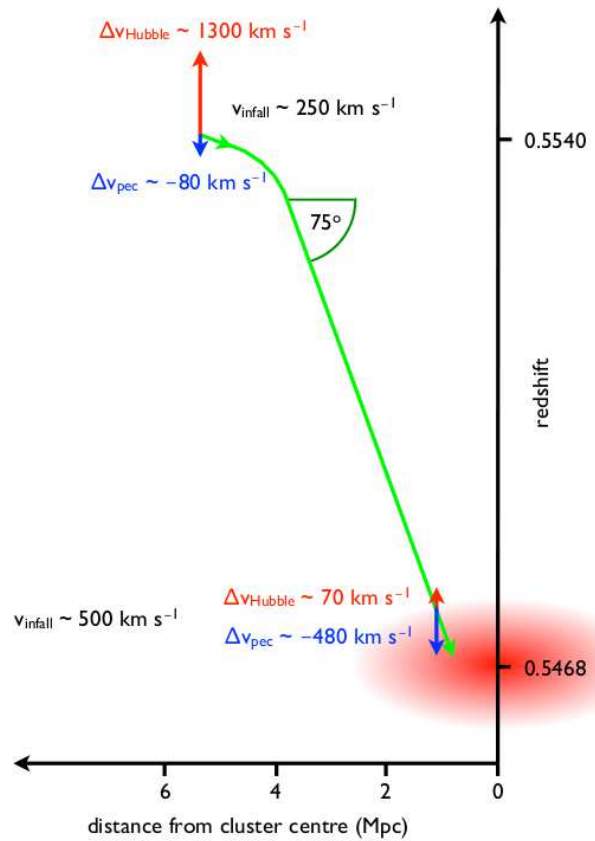}
\caption{Schematic presentation of the geometry of the large-scale filament along the line of sight, as deduced from radial-velocity measurements. Over the first 4 Mpc (in projection) the filament's average inclination angle is found to be $\alpha \sim 75¡$. Beyond this projected distance, the filament curves increasingly toward the plane of the sky, eventually rendering its projected mass surface density too low to be detectable. In this sketch, radial velocities as observed are split into two components: peculiar velocities and Hubble flow, all expressed relative to the radial velocity of the cluster core. (Adapted from Ebeling et al.\ 2012.)}
\label{fig:schema3D}
\end{figure}

To physically characterise the large-scale filament, we 1) analyse its density profile (Fig.~\ref{fig:densprof_fit}); 2) study the variation of its diameter as a function of distance to the cluster core (Fig.~\ref{fig:r_filament}); 3) derive a schematic picture of its geometry and orientation along the line of sight (Figs.~\ref{fig:schema3D} \& \ref{fig:alpha_rho}); and 4) attempt to measure the mass in stars within the entire structure (Fig.~\ref{fig:mstell}).

%______________________________________________________ 
% 6.1
%______________________________________________________
\subsection{Density Profiles}
\label{sect:densprof}
Figure~\ref{fig:densprof_fit} shows the density profile across the entire study area as a function of distance from the cluster centre as defined and shown in Figs.~\ref{fig:potgrid}, \ref{fig:kappamap}, \ref{fig:contours_slwl}, and \ref{fig:massmap_WL}. Also shown are the profiles of the filament only (obtained by only considering the WL mass along the magenta line and within the bold white contour in Fig.~\ref{fig:massmap_WL}), and of the cluster. The cluster density profile is cut at 2~$h_{74}^{-1}$Mpc from the cluster core as further, the ACS field is dominated by the filament.

As is apparent from both Figs.~\ref{fig:massmap_WL} and \ref{fig:densprof_fit}, the filament connects to the cluster at a distance of about 1.5$h_{74}^{-1}$~Mpc from the cluster core, causing the density profile to flatten from this distance outward. The mean density in this region is $(2.92 \pm 0.66)\times 10^{8} \,h_{74}$ M$_{\odot}$ kpc$^{-2}$. The embedded galaxy group detected also in the cluster light and at X-ray wavelengths is responsible for the peak in the density profile at a distance of about 2$h_{74}^{-1}$~Mpc from the cluster centre. 
The dip in the filament profile at $\sim$2.8 $h_{74}^{-1}$~Mpc from the cluster centre coincides with a dip visible also in projection in Fig.~\ref{fig:massmap_WL} (see Sect.~\ref{sect:sec63} for more details on the 3D geometry of the large-scale filament). The filament clearly dominates the radial density profile from $\sim$2 $h_{74}^{-1}$~Mpc from the cluster centre until the edge of the HST/ACS field.

%______________________________________________________
% Fig 12
%______________________________________________________
%%%%%%%%%%%%%%%%%%%%%%%%%%%%%%%%%%%%%%

%______________________________________________________
% Fig 13
%______________________________________________________
%%%%%%%%%%%%%%%%%%%%%%%%%%%%%%%%%%%%%%

Fitting any of these profiles with parametric models (NFW, SIS, etc) is not meaningful in view of the complexity and obvious non-sphericity of these structures. We note, however, that the overall density profile decreases as $R^{-2}$ until about $2\,h_{74}^{-1}$~Mpc, where the beginning of the filament causes a dramatic flattening of the density profile. For illustrative purposes, Fig.~\ref{fig:densprof_fit} shows two different slopes, $R^{-1}$ and $R^{-2}$.

%______________________________________________________ 
% 6.2
%______________________________________________________
\subsection{Filament Size}\label{sect:filsize}
We define the region of the filament on the mass map by a density threshold of $1.84 \times 10^{8}\, h_{74}$ M$_{\odot}$.kpc$^{-2}$ which corresponds to the 3~sigma detection contour of the lensing mass reconstruction, and then define the central axis of the filament with respect to this contour (magenta line in Fig.~\ref{fig:massmap_WL}). Under the simplifying assumption that the filament cross-section is spherical (i.e., the filament resembles a cylinder of variable radius), the filament diameter is given by the perpendicular distance of the contour to this axis. The uncertainty of this measurement is assessed by repeated this procedure for every mass map produced by LENSTOOL, and adopting the standard deviation of the results as the error of the filament diameter.

Figure~\ref{fig:r_filament} shows the results thus obtained as a function of distance from the cluster centre. We find the filament's diameter to decrease with increasing distance from the cluster centre, albeit with mild local variations.  Recognizable features are the local peak at 1.85 $h_{74}^{-1}$~Mpc from the cluster centre where the presence of an embedded group of galaxies (marked by a blue cross in Fig.~\ref{fig:massmap_WL}) combined with a second, X-ray faint mass concentration causes the filament to widen to $(2.79 \pm 0.45)\, h_{74}^{-1}$~Mpc. The following narrowing of the filament at a cluster-centric distance of about $2.8\,h_{74}^{-1}$~Mpc coincides with a dip in the filament density (cf.\ Sect.~\ref{sect:densprof}).
Two additional increases in size (neither of them detected in X-rays) are observed at larger distances of approximately 3.6 and $4.7\,h_{74}^{-1}$~Mpc and can again be matched to local peaks in the filament density and projected WL mass density (Figs.~\ref{fig:densprof_fit} and \ref{fig:massmap_WL}, respectively).

%______________________________________________________ 
% 6.3
%______________________________________________________
\subsection{3D-properties of the large-scale filament}
\label{sect:sec63}
In order to derive intrinsic properties of the filament from the observables measured in projection, we need
to know the three-dimensional orientation and geometry of the filament and, specifically, its inclination with
respect to the plane of the sky. Ebeling et al.\ (2012, in preparation) derive a model of the entire complex by comparing the measured variation in the mean radial velocity of galaxies along the filament axis to expectations from Hubble-flow velocities, as well as with the predictions of peculiar velocities within filaments from numerical simulations \citep{colberg05,cuesta08,ceccarelli11}.  A self-consistent description is obtained for  an average inclination angle of the filament with respect to the plane of the sky of about 75 degrees (Fig.~\ref{fig:schema3D}). At this steep inclination, the three-dimensional length of the filament reaches 18 h$^{-1}$ Mpc within our study region.

Using the weak-lensing mass map and the above model of the 3D geometry and orientation of the filament, we can constrain the filament's  intrinsic density. Adopting again a cylindrical geometry and the average projected values from Figs.~\ref{fig:densprof_fit} and \ref{fig:r_filament} we obtain a mass density of $\rho_{\rm filament} = (3.13 \pm 0.71) \times 10^{13}\,h_{74}^{2}$ M$_{\odot}$ Mpc$^{-3}$, or $$ \rho_{\rm filament} = (206 \pm 46)\,\rho_{\rm crit} $$ in units of the critical density of the Universe. 

\begin{figure}
\includegraphics[width=84mm]{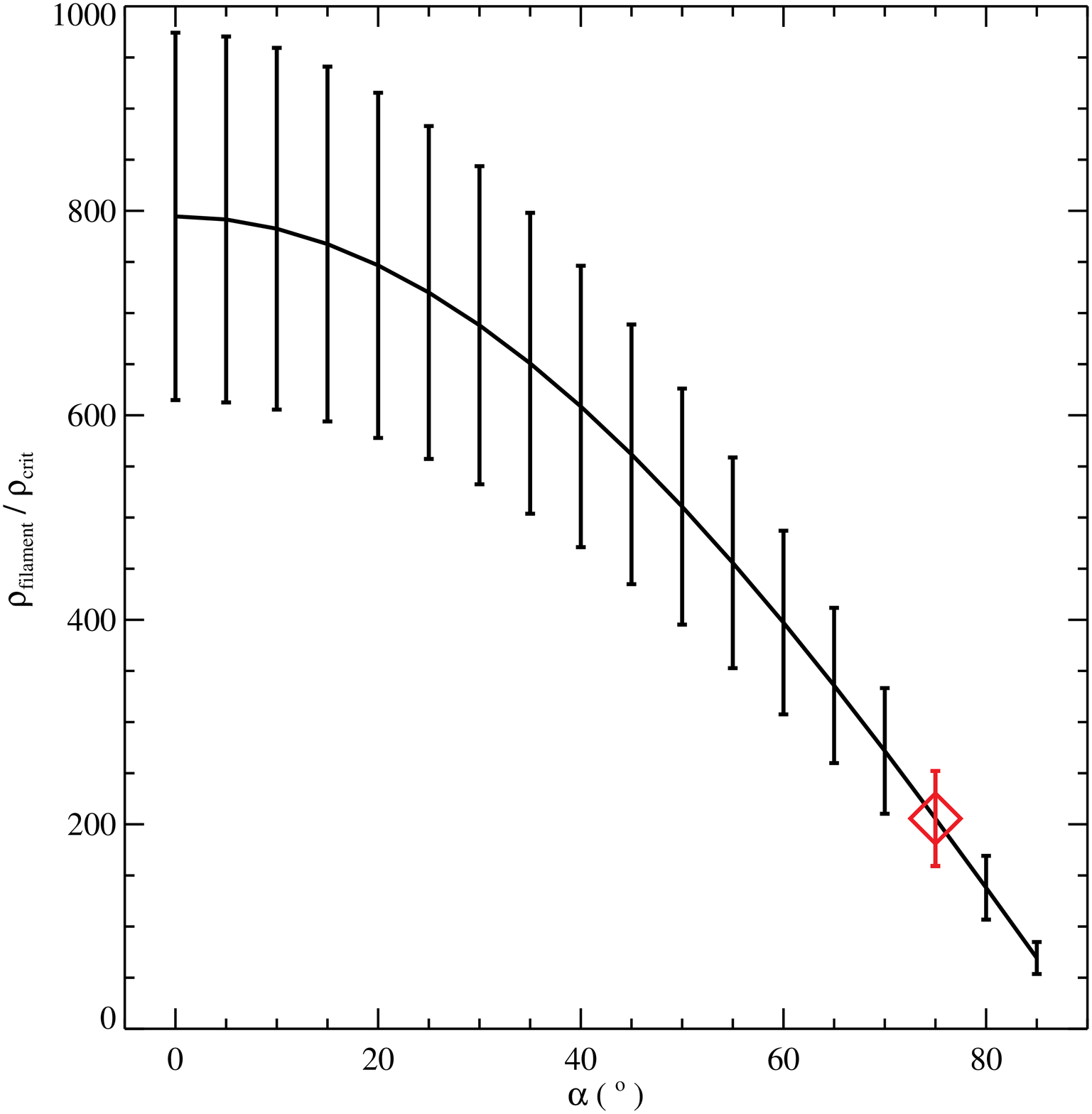}
\caption{Mass density of the filament in units of $\rho_{\rm crit}$ as a function of the orientation angle $\alpha$. The red diamond highlights the density for the adopted average inclination angle of 75\,$\deg$.}
\label{fig:alpha_rho}
\end{figure}

We stress that this is an average value: a more complex density distribution along the filament is not only possibly but also physically plausible. The density of the filament depends strongly on its inclination angle $\alpha$ which is reasonably well constrained on average but is uncertain at large cluster-centric distances (see Fig.~\ref{fig:schema3D}). Figure~\ref{fig:alpha_rho} shows how sensitively the deduced density of the filament depends on $\alpha$. In addition, embedded mass concentrations like the X-ray bright group of galaxies highlighted in Fig.~\ref{fig:massmap_WL} and clearly detected also in the mass surface density profile (Fig.~\ref{fig:densprof_fit}) will cause local variations in the filament's density.  A more comprehensive discussion of dynamical considerations regarding the 3D geometry of the cluster-filament complex is provided in Ebeling et al.\ (in preparation).

%______________________________________________________
% Fig 14
%______________________________________________________
%%%%%%%%%%%%%%%%%%%%%%%%%%%%%%%%%%%%%%

%______________________________________________________ 
% 6.4
%______________________________________________________
\subsection{Stellar Mass Fraction}
\begin{figure}
\includegraphics[width=84mm]{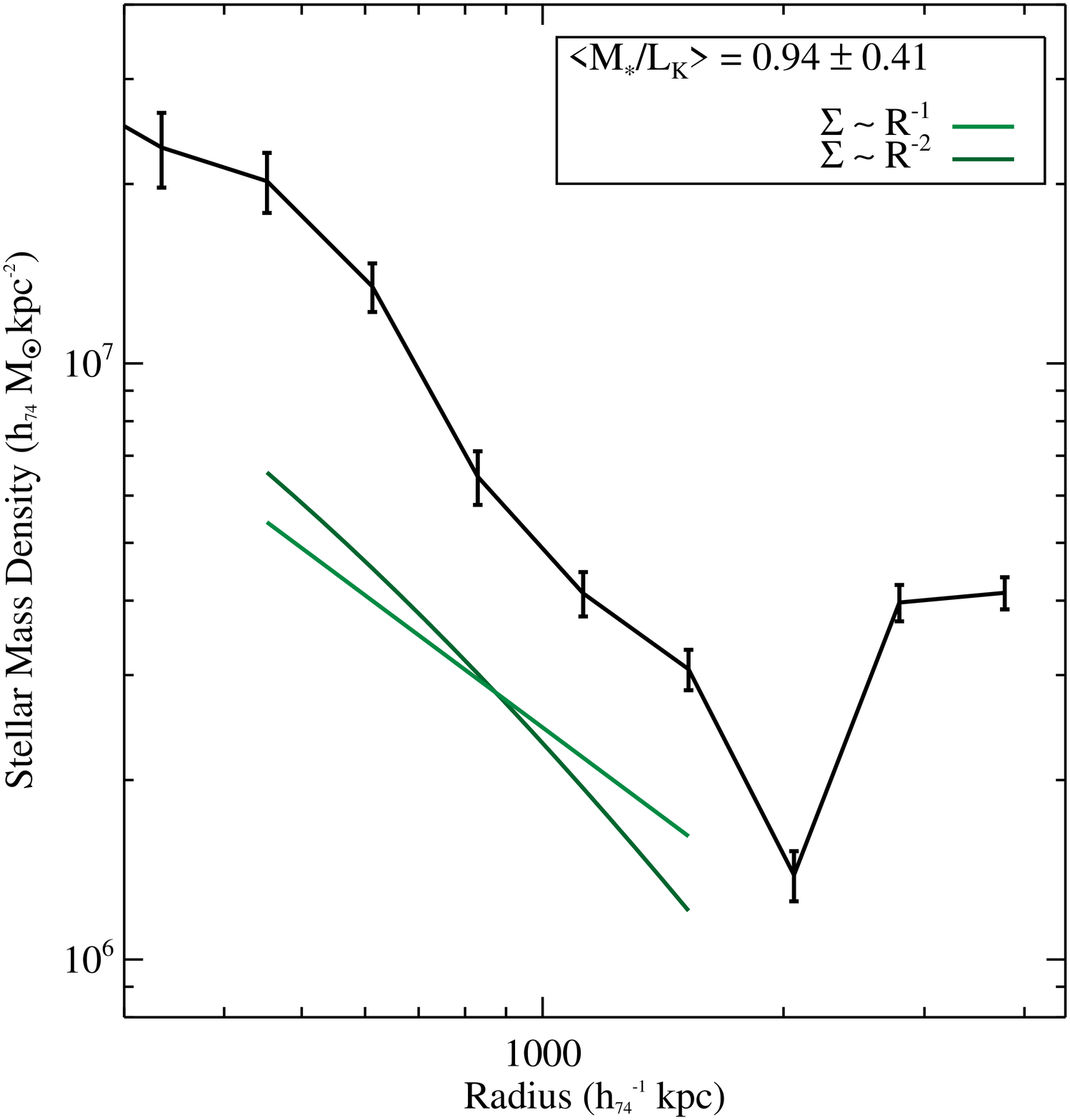}
\caption{Density profile of the stellar component in the whole HST/ACS field of MACSJ0717.5+3745. The stellar mass is obtained using the Arnouts et al. (2007) relation for a red galaxy population, and an Salpeter IMF. The mean $M_{\ast}/L_{K}$ ratio is equal to $0.94 \pm 0.41$ for our galaxy population.
The two green curves represent two different slopes: $R^{-1}$ $\&$ $R^{-2}$, light and dark green respectively.}
\label{fig:mstell}
\end{figure}

To compute the stellar mass, $M_\ast$,  across our study region we use the relation $\log (M_\ast/L_{\rm K}) = az + b$ established by \citet{arnouts07} for quiescent (red) galaxies in the the VVDS sample \citep{lefevre05} and adopting a Salpeter initial mass function (IMF). Here $L_{\rm K}$ is the galaxy's luminosity in the K band, $z$ is its redshift, and the parameters $a$ and $b$ are given by
$$a = -0.18 \pm 0.03 \quad \quad
b = -0.05 \pm 0.03 .
$$
We apply this relation to the K-band magnitudes of all cluster members, defined using the redshift criteria presented in Sect.~\ref{sect:sec3}. The K-band magnitude limit of our catalogue is 23.1. The resulting projected mass density in stars is shown in Fig.~\ref{fig:mstell} as a function of  distance from the cluster centre. It decreases in proportion
to the total projected mass density depicted in Fig.~\ref{fig:densprof_fit}, as illustrated by the two different slopes, $\Sigma \sim R^{-1}$ and $\Sigma \sim R^{-2}$, shown in both figures. Hence the measured mass-to-light ratio is approximately constant across the study area, with a mean value of $\langle M_{\ast}/L_{K}\rangle=0.94 \pm 0.41$. 

%______________________________________________________
% Fig 15
%______________________________________________________
%%%%%%%%%%%%%%%%%%%%%%%%%%%%%%%%%%%%%%

To compare our result with those obtained by \cite{leauthaud11}  for COSMOS data we need to adjust our measurement to account for the different IMF used by these authors. Applying a shift of 0.25 dex to our masses to convert from a Salpeter IMF to a Chabrier IMF, we find $\langle M_{\ast}/L_{K}\rangle_{\rm Chabrier} = 0.73 \pm 0.22$ for quiescent galaxies at $z\sim 0.5$, in good agreement with \cite{leauthaud11} .

The fraction of the total mass in stars, $f_{\ast}$, i.e., the ratio between the stellar mass and the total mass  of the cluster derived from our weak-lensing analysis, is 
$f_{\ast} = (1.3 \pm 0.4)\%$ and $f_{\ast} = (0.9 \pm 0.2) \%$ for a Salpeter and a Chabrier IMF, respectively. The latter value is slightly lower than that obtained for COSMOS data by \cite{leauthaud11}, possibly because of different limiting K-band magnitudes or differences in the galaxy environments probed (the COSMOS study was conducted for groups with a halo mass comprised between $10^{11}$ and $10^{14}$~M$_{\odot}$ and extrapolated to halos of $\sim 10^{15}$ $h^{-1}$ M$_{\odot}$).
\\
Finally, we compute the total stellar mass within our study area and find $M_{\ast} = (8.62 \pm 0.24)\times 10^{13}$~M$_{\odot}$ and $M_{\ast} = (6.71 \pm 0.19)\times 10^{13}$~M$_{\odot}$ for a Salpeter and a Chabrier IMF, respectively.

%______________________________________________________ 
% SECTION 7:  SUMMARY & DISCUSSION
%______________________________________________________
\section{Summary and Discussion}
We present the results of a weak-gravitational lensing analysis of the massive galaxy cluster MACSJ0717.5+3745 and its large-scale filament, based on a mosaic composed of 18 HST/ACS images covering an area of approximately $10\times 20$ arcmin$^2$. Our mass reconstruction method uses RRG shape measurements \citep{rhodes07}; a multi-scale adaptive grid designed to follow the structures' K-band light and including galaxy-size potentials to account for cluster members; and the LENSTOOL software package, improved by the implementation of a Bayesian MCMC optimisation method that allows the propagation of measurement uncertainties into errors on the filament mass. As a critical step of the analysis, we use spectroscopic and photometric redshifts, as well as colour-colour cuts, all based on groundbased observations, to eliminate foreground galaxies and cluster members, thereby reducing dilution of the shear signal from unlensed galaxies.

A simple convergence map of the study area, obtained with the inversion method of \cite{seitz95}, already allows the detection of the cluster core (at more than $6\sigma$ significance) and of two extended mass concentrations (at 2 to 3$\sigma$ significance) near the beginning and (apparent) end of the filament. 

The fully optimised weak-lensing mass model yields the  surface mass density shown in Fig.~\ref{fig:massmap_WL}. Its validity is confirmed by the excellent agreement between the mass of the cluster core measured by us, $M_{\rm WL} (R<500 \,{\rm kpc}) = (1.04 \pm 0.08)\times 10^{15} \,h_{74}^{-1}$ M$_{\odot}$, and the one obtained in a strong-lensing analysis by \cite{limousin11}, $M_{\rm SL} (R<500 \,{\rm kpc}) = (1.06 \pm 0.03)\times 10^{15}\,h_{74}^{-1}$ M$_{\odot}$.

Based on our weak-lensing mass reconstruction, we report the first unambiguous detection of a large-scale filament fueling the growth of a massive galaxy cluster at a node of the Cosmic Web.  
The projected length of the filament is approximately $4.5\,h_{74}^{-1}$ Mpc, and its mean mass surface density  $(2.92 \pm 0.66) \times 10^{8}\, h_{74}$  M$_{\odot}$ kpc$^{-2}$. We measure the width and mass surface density of the filament as a function of distance from the cluster centre, and find both to decrease, albeit with local variations due to at least four embedded mild mass concentrations. One of these, at a projected distance of 1.85\,$h_{74}^{-1}$ Mpc from the cluster centre, coincides with an X-ray detected galaxy group. The filament is found to narrow at a cluster-centric distance of approximately $3.6\,h_{74}^{-1}$ Mpc as it curves south and, most likely, also recedes from us. We find the cluster's mass surface density to decrease as $r^{-2}$ until the onset of the filament flattens the profile dramatically.

Following the analysis by Ebeling et al.\ (2012, in preparation) we adopt an average inclination angle with respect to the plane of the sky of 75$^\circ$ for the majority of the filament's length. For this scenario and under the simplifying assumption of a cylindrical cross-section we obtain estimates of the filament's three-dimensional length and average mass density in units of the Universe's critical density of $18\,h^{-1}_{74}$ Mpc and $(206 \pm 46)\,\rho_{\rm crit}$, respectively. However,  additional systematic uncertainties enter since either quantity is sensitive to the adopted inclination angle. These values lie at the high end of the range predicted from numerical simulations \citep[e.g.,][]{colberg05}, which is not unexpected given the extreme mass of MACSJ0717.5+3745.

The lensing surface mass density and the spectroscopic redshift distribution suggest the  consistent picture of an elongated structure at the redshift of the cluster. The galaxy distribution along the filament appears to be homogeneous, and at the cluster redshift \citep[see][]{ebeling04}. This motivates our conclusion of an unambiguous detection of a large scale filament, and not a superposition of galaxy groups projected on the plane of the sky.

Finally, we measure the stellar mass fraction in the entire MACSJ0717.5+3745 field, using as a proxy the K-band luminosity of galaxies with redshifts consistent with that of the cluster-filament complex. We find $f_{\ast} = (1.3 \pm 0.4)\%$ and $f_{\ast} = (0.9 \pm 0.2) \%$ for a Salpeter and Chabrier IMF, respectively, in good agreement with previous results in the fields of massive clusters \citep{leauthaud11}.

Our results show that, if shear dilution by unlensed galaxies can be efficiently suppressed, weak-lensing studies of massive clusters are capable of detecting and mapping the complex mass distribution at the vertices of the cosmic web. Confirming results of numerical simulations \citep[e.g.][]{colberg99, colberg05,OH11}, our weak-lensing mass reconstruction shows that the contribution from large-scale filaments can be significant and needs to be taken into account in the modelling of mass density profiles. Expanding this kind of investigation to HST/ACS-based weak-lensing studies of other MACS clusters will allow us to constrain the properties of large-scale filaments and the dynamics of cluster growth on a sound statistical basis.

%______________________________________________________ 
% ACKNOLEDGMENTS
%______________________________________________________
\section*{Acknowledgments}
We thank Douglas Clowe for accepting to review this paper and for his useful comments. MJ would like to thank Graham P.\ Smith, Kavilan Moodley, and, Pierre-Yves Chabaud for useful discussions. EJ acknowledges support from the Jet Propulsion Laboratory under contract with the California Institute of Technology, the NASA Postdoctoral Program, and, the Centre National d'Etudes Spatiales.
MJ and EJ are indebted to Jason Rhodes for discussions and advice. HE gratefully acknowledges financial support from STScI grant GO-10420. We thank the UH Time Allocation Committee for their support of the extensive groundbased follow-up observations required for this study.
ML acknowledges the Centre National de la Recherche Scientifique (CNRS) for its support. The Dark Cosmology Centre is funded by the Danish National Research Foundation. This work was performed using facilities offered by CeSAM (Centre de donn\'eeS Astrophysique de Marseille-(http://lam.oamp.fr/cesam/).
This work was supported by World Premier International Research Center Initiative (WPI Initiative), MEXT, Japan. 

\bibliographystyle{mn2e}
\bibliography{reference}

\begin{thebibliography}{}

\bibitem[\protect\citeauthoryear{{Arag{\'o}n-Calvo}, {van de Weygaert}, {Jones}
  \& {van der Hulst}}{{Arag{\'o}n-Calvo} et~al.}{2007}]{aragoncalvo07}
{Arag{\'o}n-Calvo} M.~A.,  {van de Weygaert} R.,  {Jones} B.~J.~T.,    {van der
  Hulst} J.~M.,  2007, \apjl, 655, L5

\bibitem[\protect\citeauthoryear{{Arnouts}, {Cristiani}, {Moscardini},
  {Matarrese}, {Lucchin}, {Fontana} \& {Giallongo}}{{Arnouts}
  et~al.}{1999}]{arnouts99}
{Arnouts} S.,  {Cristiani} S.,  {Moscardini} L.,  {Matarrese} S.,  {Lucchin}
  F.,  {Fontana} A.,    {Giallongo} E.,  1999, \mnras, 310, 540

\bibitem[\protect\citeauthoryear{{Arnouts}, {Walcher}, {Le F{\`e}vre},
  {Zamorani}, {Ilbert}, {Le Brun}, {Pozzetti}, {Bardelli}, {Tresse}, {Zucca},
  {Charlot} \& [...]}{{Arnouts} et~al.}{2007}]{arnouts07}
{Arnouts} S.,  {Walcher} C.~J.,  {Le F{\`e}vre} O.,  {Zamorani} G.,  {Ilbert}
  O.,  {Le Brun} V.,  {Pozzetti} L.,  {Bardelli} S.,  {Tresse} L.,  {Zucca} E.,
   {Charlot} S.,    [...] 2007, \aap, 476, 137

\bibitem[\protect\citeauthoryear{{Bertin} \& {Arnouts}}{{Bertin} \&
  {Arnouts}}{1996}]{BA96}
{Bertin} E.,  {Arnouts} S.,  1996, \aap, 117, 393

\bibitem[\protect\citeauthoryear{{Bond}, {Kofman} \& {Pogosyan}}{{Bond}
  et~al.}{1996}]{bond96}
{Bond} J.~R.,  {Kofman} L.,    {Pogosyan} D.,  1996, \nat, 380, 603

\bibitem[\protect\citeauthoryear{{Bond}, {Strauss} \& {Cen}}{{Bond}
  et~al.}{2010}]{bond10}
{Bond} N.~A.,  {Strauss} M.~A.,    {Cen} R.,  2010, \mnras, 409, 156

\bibitem[\protect\citeauthoryear{{Casertano}, {de Mello}, {Dickinson},
  {Ferguson}, {Fruchter}, {Gonzalez-Lopezlira}, {Heyer}, {Hook}, {Levay},
  {Lucas}, {Mack}, {Makidon}, {Mutchler}, {Smith}, {Stiavelli}, {Wiggs} \&
  {Williams}}{{Casertano} et~al.}{2000}]{casertano00}
{Casertano} S.,  {de Mello} D.,  {Dickinson} M.,  {Ferguson} H.~C.,  {Fruchter}
  A.~S.,  {Gonzalez-Lopezlira} R.~A.,  {Heyer} I.,  {Hook} R.~N.,  {Levay} Z.,
  {Lucas} R.~A.,  {Mack} J.,  {Makidon} R.~B.,  {Mutchler} M.,  {Smith} T.~E.,
  {Stiavelli} M.,  {Wiggs} M.~S.,    {Williams} R.~E.,  2000, \aj, 120, 2747

\bibitem[\protect\citeauthoryear{{Ceccarelli}, {Paz}, {Padilla} \&
  {Lambas}}{{Ceccarelli} et~al.}{2011}]{ceccarelli11}
{Ceccarelli} L.,  {Paz} D.~J.,  {Padilla} N.,    {Lambas} D.~G.,  2011, \mnras,
  412, 1778

\bibitem[\protect\citeauthoryear{{Cen} \& {Ostriker}}{{Cen} \&
  {Ostriker}}{1999}]{CO99}
{Cen} R.,  {Ostriker} J.~P.,  1999, \apjl, 519, L109

\bibitem[\protect\citeauthoryear{{Clowe}, {Luppino}, {Kaiser}, {Henry} \&
  {Gioia}}{{Clowe} et~al.}{1998}]{clowe98}
{Clowe} D.,  {Luppino} G.~A.,  {Kaiser} N.,  {Henry} J.~P.,    {Gioia} I.~M.,
  1998, \apjl, 497, L61+

\bibitem[\protect\citeauthoryear{{Colberg}, {Krughoff} \& {Connolly}}{{Colberg}
  et~al.}{2005}]{colberg05}
{Colberg} J.~M.,  {Krughoff} K.~S.,    {Connolly} A.~J.,  2005, \mnras, 359,
  272

\bibitem[\protect\citeauthoryear{{Colberg}, {White}, {Jenkins} \&
  {Pearce}}{{Colberg} et~al.}{1999}]{colberg99}
{Colberg} J.~M.,  {White} S.~D.~M.,  {Jenkins} A.,    {Pearce} F.~R.,  1999,
  \mnras, 308, 593

\bibitem[\protect\citeauthoryear{{Colless}, {Dalton}, {Maddox}, {Sutherland},
  {Norberg}, {Cole}, {Bland-Hawthorn}, {Bridges}, {Cannon}, {Collins}, {Couch},
  {Cross} \& [...]}{{Colless} et~al.}{2001}]{colless01}
{Colless} M.,  {Dalton} G.,  {Maddox} S.,  {Sutherland} W.,  {Norberg} P.,
  {Cole} S.,  {Bland-Hawthorn} J.,  {Bridges} T.,  {Cannon} R.,  {Collins} C.,
  {Couch} W.,  {Cross} N.,    [...] 2001, \mnras, 328, 1039

\bibitem[\protect\citeauthoryear{{Cuesta}, {Prada}, {Klypin} \&
  {Moles}}{{Cuesta} et~al.}{2008}]{cuesta08}
{Cuesta} A.~J.,  {Prada} F.,  {Klypin} A.,    {Moles} M.,  2008, \mnras, 389,
  385

\bibitem[\protect\citeauthoryear{{Diego}, {Tegmark}, {Protopapas} \&
  {Sandvik}}{{Diego} et~al.}{2007}]{diego07}
{Diego} J.~M.,  {Tegmark} M.,  {Protopapas} P.,    {Sandvik} H.~B.,  2007,
  \mnras, 375, 958

\bibitem[\protect\citeauthoryear{{Dietrich}, {Schneider}, {Clowe},
  {Romano-D{\'{\i}}az} \& {Kerp}}{{Dietrich} et~al.}{2005}]{dietrich05}
{Dietrich} J.~P.,  {Schneider} P.,  {Clowe} D.,  {Romano-D{\'{\i}}az} E.,
  {Kerp} J.,  2005, \aap, 440, 453

\bibitem[\protect\citeauthoryear{{Donovan}}{{Donovan}}{2007}]{donovan07}
{Donovan} D.~A.~K.,  2007, PhD thesis, University of Hawai'i at Manoa

\bibitem[\protect\citeauthoryear{{Ebeling}, {Barrett} \& {Donovan}}{{Ebeling}
  et~al.}{2004}]{ebeling04}
{Ebeling} H.,  {Barrett} E.,    {Donovan} D.,  2004, \apjl, 609, L49

\bibitem[\protect\citeauthoryear{{Ebeling}, {Barrett}, {Donovan}, {Ma}, {Edge}
  \& {van Speybroeck}}{{Ebeling} et~al.}{2007}]{ebeling07}
{Ebeling} H.,  {Barrett} E.,  {Donovan} D.,  {Ma} C.-J.,  {Edge} A.~C.,    {van
  Speybroeck} L.,  2007, \apjl, 661, L33

\bibitem[\protect\citeauthoryear{{Ebeling}, {Edge} \& {Henry}}{{Ebeling}
  et~al.}{2001}]{ebeling01}
{Ebeling} H.,  {Edge} A.~C.,    {Henry} J.~P.,  2001, \apj, 553, 668

\bibitem[\protect\citeauthoryear{{Edge}, {Ebeling}, {Bremer}, {R{\"o}ttgering},
  {van Haarlem}, {Rengelink} \& {Courtney}}{{Edge} et~al.}{2003}]{edge03}
{Edge} A.~C.,  {Ebeling} H.,  {Bremer} M.,  {R{\"o}ttgering} H.,  {van Haarlem}
  M.~P.,  {Rengelink} R.,    {Courtney} N.~J.~D.,  2003, \mnras, 339, 913

\bibitem[\protect\citeauthoryear{{El{\'{\i}}asd{\'o}ttir}, {Limousin},
  {Richard}, {Hjorth}, {Kneib}, {Natarajan}, {Pedersen}, {Jullo} \&
  {Paraficz}}{{El{\'{\i}}asd{\'o}ttir} et~al.}{2007}]{eliasdottir07}
{El{\'{\i}}asd{\'o}ttir} {\'A}.,  {Limousin} M.,  {Richard} J.,  {Hjorth} J.,
  {Kneib} J.-P.,  {Natarajan} P.,  {Pedersen} K.,  {Jullo} E.,    {Paraficz}
  D.,  2007, ArXiv e-prints, 710

\bibitem[\protect\citeauthoryear{{Faber} \& {Jackson}}{{Faber} \&
  {Jackson}}{1976}]{FJ76}
{Faber} S.~M.,  {Jackson} R.~E.,  1976, \apj, 204, 668

\bibitem[\protect\citeauthoryear{{Fang}, {Canizares} \& {Yao}}{{Fang}
  et~al.}{2007}]{fang07}
{Fang} T.,  {Canizares} C.~R.,    {Yao} Y.,  2007, \apj, 670, 992

\bibitem[\protect\citeauthoryear{{Fang}, {Marshall}, {Lee}, {Davis} \&
  {Canizares}}{{Fang} et~al.}{2002}]{fang02}
{Fang} T.,  {Marshall} H.~L.,  {Lee} J.~C.,  {Davis} D.~S.,    {Canizares}
  C.~R.,  2002, \apjl, 572, L127

\bibitem[\protect\citeauthoryear{{Fritz}, {Ziegler}, {Bower}, {Smail} \&
  {Davies}}{{Fritz} et~al.}{2005}]{fritz05}
{Fritz} A.,  {Ziegler} B.~L.,  {Bower} R.~G.,  {Smail} I.,    {Davies} R.~L.,
  2005, \mnras, 358, 233

\bibitem[\protect\citeauthoryear{{Galeazzi}, {Gupta} \& {Ursino}}{{Galeazzi}
  et~al.}{2009}]{galeazzi09}
{Galeazzi} M.,  {Gupta} A.,    {Ursino} E.,  2009, \apj, 695, 1127

\bibitem[\protect\citeauthoryear{{Gavazzi}, {Mellier}, {Fort}, {Cuillandre} \&
  {Dantel-Fort}}{{Gavazzi} et~al.}{2004}]{gavazzi04}
{Gavazzi} R.,  {Mellier} Y.,  {Fort} B.,  {Cuillandre} J.-C.,    {Dantel-Fort}
  M.,  2004, \aap, 422, 407

\bibitem[\protect\citeauthoryear{{Geller} \& {Huchra}}{{Geller} \&
  {Huchra}}{1989}]{GH89}
{Geller} M.~J.,  {Huchra} J.~P.,  1989, Science, 246, 897

\bibitem[\protect\citeauthoryear{{Gray}, {Taylor}, {Meisenheimer}, {Dye},
  {Wolf} \& {Thommes}}{{Gray} et~al.}{2002}]{gray02}
{Gray} M.~E.,  {Taylor} A.~N.,  {Meisenheimer} K.,  {Dye} S.,  {Wolf} C.,
  {Thommes} E.,  2002, \apj, 568, 141

\bibitem[\protect\citeauthoryear{{Hahn}, {Carollo}, {Porciani} \&
  {Dekel}}{{Hahn} et~al.}{2007}]{hahn07}
{Hahn} O.,  {Carollo} C.~M.,  {Porciani} C.,    {Dekel} A.,  2007, \mnras, 381,
  41

\bibitem[\protect\citeauthoryear{{Heymans}, {Gray}, {Peng}, {van Waerbeke},
  {Bell}, {Wolf}, {Bacon}, {Balogh}, {Barazza}, {Barden}, {B{\"o}hm},
  {Caldwell} \& [É]}{{Heymans} et~al.}{2008}]{heymans08}
{Heymans} C.,  {Gray} M.~E.,  {Peng} C.~Y.,  {van Waerbeke} L.,  {Bell} E.~F.,
  {Wolf} C.,  {Bacon} D.,  {Balogh} M.,  {Barazza} F.~D.,  {Barden} M.,
  {B{\"o}hm} A.,  {Caldwell}   [É] 2008, \mnras, 385, 1431

\bibitem[\protect\citeauthoryear{{Heymans}, {Van Waerbeke}, {Bacon}, {Berge},
  {Bernstein}, {Bertin}, {Bridle}, {Brown}, {Clowe}, {Dahle}, {Erben}, {Gray}
  \& [...]}{{Heymans} et~al.}{2006}]{heymans06}
{Heymans} C.,  {Van Waerbeke} L.,  {Bacon} D.,  {Berge} J.,  {Bernstein} G.,
  {Bertin} E.,  {Bridle} S.,  {Brown} M.~L.,  {Clowe} D.,  {Dahle} H.,  {Erben}
  T.,  {Gray} M.,    [...] 2006, \mnras, 368, 1323

\bibitem[\protect\citeauthoryear{{Huchra} \& {Geller}}{{Huchra} \&
  {Geller}}{1982}]{HG82}
{Huchra} J.~P.,  {Geller} M.~J.,  1982, \apj, 257, 423

\bibitem[\protect\citeauthoryear{{Ilbert}, {Arnouts}, {McCracken},
  {Bolzonella}, {Bertin}, {Le F{\`e}vre}, {Mellier}, {Zamorani}, {Pell{\`o}},
  {Iovino}, {Tresse}, {Le Brun} \& [...]}{{Ilbert} et~al.}{2006}]{ilbert06}
{Ilbert} O.,  {Arnouts} S.,  {McCracken} H.~J.,  {Bolzonella} M.,  {Bertin} E.,
   {Le F{\`e}vre} O.,  {Mellier} Y.,  {Zamorani} G.,  {Pell{\`o}} R.,  {Iovino}
  A.,  {Tresse} L.,  {Le Brun} V.,    [...] 2006, \aap, 457, 841

\bibitem[\protect\citeauthoryear{{Ilbert}, {Capak}, {Salvato}, {Aussel},
  {McCracken}, {Sanders}, {Scoville}, {Kartaltepe}, {Arnouts}, {Le Floc'h},
  {Mobasher} \& [...]}{{Ilbert} et~al.}{2009}]{ilbert09}
{Ilbert} O.,  {Capak} P.,  {Salvato} M.,  {Aussel} H.,  {McCracken} H.~J.,
  {Sanders} D.~B.,  {Scoville} N.,  {Kartaltepe} J.,  {Arnouts} S.,  {Le
  Floc'h} E.,  {Mobasher} B.,    [...] 2009, \apj, 690, 1236

\bibitem[\protect\citeauthoryear{{Jullo} \& {Kneib}}{{Jullo} \&
  {Kneib}}{2009}]{jullo09}
{Jullo} E.,  {Kneib} J.,  2009, \mnras, 395, 1319

\bibitem[\protect\citeauthoryear{{Jullo}, {Kneib}, {Limousin},
  {El{\'{\i}}asd{\'o}ttir}, {Marshall} \& {Verdugo}}{{Jullo}
  et~al.}{2007}]{jullo07}
{Jullo} E.,  {Kneib} J.-P.,  {Limousin} M.,  {El{\'{\i}}asd{\'o}ttir} {\'A}.,
  {Marshall} P.~J.,    {Verdugo} T.,  2007, New Journal of Physics, 9, 447

\bibitem[\protect\citeauthoryear{{Kaastra}, {Werner}, {Herder}, {Paerels}, {de
  Plaa}, {Rasmussen} \& {de Vries}}{{Kaastra} et~al.}{2006}]{kaastra06}
{Kaastra} J.~S.,  {Werner} N.,  {Herder} J.~W.~A.~d.,  {Paerels} F.~B.~S.,  {de
  Plaa} J.,  {Rasmussen} A.~P.,    {de Vries} C.~P.,  2006, \apj, 652, 189

\bibitem[\protect\citeauthoryear{{Kaiser}, {Wilson}, {Luppino}, {Kofman},
  {Gioia}, {Metzger} \& {Dahle}}{{Kaiser} et~al.}{1998}]{kaiser98}
{Kaiser} N.,  {Wilson} G.,  {Luppino} G.,  {Kofman} L.,  {Gioia} I.,  {Metzger}
  M.,    {Dahle} H.,  1998, \arxiv

\bibitem[\protect\citeauthoryear{{Kartaltepe}, {Ebeling}, {Ma} \&
  {Donovan}}{{Kartaltepe} et~al.}{2008}]{kartaltepe08}
{Kartaltepe} J.~S.,  {Ebeling} H.,  {Ma} C.~J.,    {Donovan} D.,  2008, \mnras,
  389, 1240

\bibitem[\protect\citeauthoryear{{Kassiola} \& {Kovner}}{{Kassiola} \&
  {Kovner}}{1993}]{kassiola93}
{Kassiola} A.,  {Kovner} I.,  1993, \apj, 417, 450

\bibitem[\protect\citeauthoryear{{Kneib}, {Ellis}, {Smail}, {Couch} \&
  {Sharples}}{{Kneib} et~al.}{1996}]{kneib96}
{Kneib} J.-P.,  {Ellis} R.~S.,  {Smail} I.,  {Couch} W.~J.,    {Sharples}
  R.~M.,  1996, \apj, 471, 643

\bibitem[\protect\citeauthoryear{{Koekemoer}, {Fruchter}, {Hook} \&
  {Hack}}{{Koekemoer} et~al.}{2002}]{koekemoer02}
{Koekemoer} A.~M.,  {Fruchter} A.~S.,  {Hook} R.~N.,    {Hack} W.,  2002, in
  {S.~Arribas, A.~Koekemoer, \& B.~Whitmore} ed., The 2002 HST Calibration
  Workshop : Hubble after the Installation of the ACS and the NICMOS Cooling
  System {MultiDrizzle: An Integrated Pyraf Script for Registering, Cleaning
  and Combining Images}.
pp 337--+

\bibitem[\protect\citeauthoryear{{Le F{\`e}vre}, {Vettolani}, {Garilli},
  {Tresse}, {Bottini}, {Le Brun}, {Maccagni}, {Picat}, {Scaramella},
  {Scodeggio} \& [É]}{{Le F{\`e}vre} et~al.}{2005}]{lefevre05}
{Le F{\`e}vre} O.,  {Vettolani} G.,  {Garilli} B.,  {Tresse} L.,  {Bottini} D.,
   {Le Brun} V.,  {Maccagni} D.,  {Picat} J.~P.,  {Scaramella} R.,  {Scodeggio}
  M.,    [É] 2005, \aap, 439, 845

\bibitem[\protect\citeauthoryear{{Leauthaud}, {Finoguenov}, {Kneib}, {Taylor},
  {Massey}, {Rhodes}, {Ilbert}, {Bundy}, {Tinker}, {George}, {Capak} \&
  [...]}{{Leauthaud} et~al.}{2010}]{leauthaud10}
{Leauthaud} A.,  {Finoguenov} A.,  {Kneib} J.,  {Taylor} J.~E.,  {Massey} R.,
  {Rhodes} J.,  {Ilbert} O.,  {Bundy} K.,  {Tinker} J.,  {George} M.~R.,
  {Capak} P.,    [...] 2010, \apj, 709, 97

\bibitem[\protect\citeauthoryear{{Leauthaud}, {Massey}, {Kneib}, {Rhodes},
  {Johnston}, {Capak}, {Heymans}, {Ellis}, {Koekemoer}, {Le F{\`e}vre},
  {Mellier} \& [...]}{{Leauthaud} et~al.}{2007}]{leauthaud07}
{Leauthaud} A.,  {Massey} R.,  {Kneib} J.,  {Rhodes} J.,  {Johnston} D.~E.,
  {Capak} P.,  {Heymans} C.,  {Ellis} R.~S.,  {Koekemoer} A.~M.,  {Le
  F{\`e}vre} O.,  {Mellier} Y.,    [...] 2007, \apjs, 172, 219

\bibitem[\protect\citeauthoryear{{Leauthaud}, {Tinker}, {Bundy}, {Behroozi},
  {Massey}, {Rhodes}, {George}, {Kneib}, {Benson}, {Wechsler}, {Busha}, {Capak}
  \& [É]}{{Leauthaud} et~al.}{2012}]{leauthaud11}
{Leauthaud} A.,  {Tinker} J.,  {Bundy} K.,  {Behroozi} P.~S.,  {Massey} R.,
  {Rhodes} J.,  {George} M.~R.,  {Kneib} J.-P.,  {Benson} A.,  {Wechsler}
  R.~H.,  {Busha} M.~T.,  {Capak} P.,    [É] 2012, \apj, 744, 159

\bibitem[\protect\citeauthoryear{{Limousin}, {Ebeling}, {Richard}, {Swinbank},
  {Smith}, {Rodionov}, {Ma}, {Smail}, {Edge}, {Jauzac}, {Jullo} \&
  {Kneib}}{{Limousin} et~al.}{2012}]{limousin11}
{Limousin} M.,  {Ebeling} H.,  {Richard} J.,  {Swinbank} A.~M.,  {Smith} G.~P.,
   {Rodionov} S.,  {Ma} C.~.,  {Smail} I.,  {Edge} A.~C.,  {Jauzac} M.,
  {Jullo} E.,    {Kneib} J.~P.,  2012, ArXiv e-prints

\bibitem[\protect\citeauthoryear{{Limousin}, {Kneib} \& {Natarajan}}{{Limousin}
  et~al.}{2005}]{limousin05}
{Limousin} M.,  {Kneib} J.-P.,    {Natarajan} P.,  2005, \mnras, 356, 309

\bibitem[\protect\citeauthoryear{{Limousin}, {Richard}, {Jullo}, {Kneib},
  {Fort}, {Soucail}, {El{\'{\i}}asd{\'o}ttir}, {Natarajan}, {Ellis}, {Smail},
  {Czoske}, {Smith}, {Hudelot}, {Bardeau}, {Ebeling}, {Egami} \&
  {Knudsen}}{{Limousin} et~al.}{2007}]{limousin07b}
{Limousin} M.,  {Richard} J.,  {Jullo} E.,  {Kneib} J.~P.,  {Fort} B.,
  {Soucail} G.,  {El{\'{\i}}asd{\'o}ttir} A.,  {Natarajan} P.,  {Ellis} R.~S.,
  {Smail} I.,  {Czoske} O.,  {Smith} G.~P.,  {Hudelot} P.,  {Bardeau} S.,
  {Ebeling} H.,  {Egami} E.,    {Knudsen} K.~K.,  2007, \apj, 668, 643

\bibitem[\protect\citeauthoryear{{Lupton}, {Gunn}, {Ivezi{\'c}}, {Knapp} \&
  {Kent}}{{Lupton} et~al.}{2001}]{lupton01}
{Lupton} R.,  {Gunn} J.~E.,  {Ivezi{\'c}} Z.,  {Knapp} G.~R.,    {Kent} S.,
  2001, in {F.~R.~Harnden Jr., F.~A.~Primini, \& H.~E.~Payne} ed., Astronomical
  Data Analysis Software and Systems X Vol.~238 of Astronomical Society of the
  Pacific Conference Series, {The SDSS Imaging Pipelines}.
pp 269--+

\bibitem[\protect\citeauthoryear{{Ma} \& {Ebeling}}{{Ma} \&
  {Ebeling}}{2010}]{ma10}
{Ma} C.,  {Ebeling} H.,  2010, \mnras, pp 1646--+

\bibitem[\protect\citeauthoryear{{Ma}, {Ebeling} \& {Barrett}}{{Ma}
  et~al.}{2009}]{ma09}
{Ma} C.-J.,  {Ebeling} H.,    {Barrett} E.,  2009, \apjl, 693, L56

\bibitem[\protect\citeauthoryear{{Ma}, {Ebeling}, {Donovan} \& {Barrett}}{{Ma}
  et~al.}{2008}]{ma08}
{Ma} C.-J.,  {Ebeling} H.,  {Donovan} D.,    {Barrett} E.,  2008, \apj, 684,
  160

\bibitem[\protect\citeauthoryear{{Massey}, {Bacon}, {Refregier} \&
  {Ellis}}{{Massey} et~al.}{2002}]{massey02}
{Massey} R.,  {Bacon} D.,  {Refregier} A.,    {Ellis} R.,  2002, in
  {N.~Metcalfe \& T.~Shanks} ed., A New Era in Cosmology Vol.~283 of
  Astronomical Society of the Pacific Conference Series, {Cosmic Shear with
  Keck: Systematic Effects}.
p.~193

\bibitem[\protect\citeauthoryear{{Massey}, {Heymans}, {Berg{\'e}}, {Bernstein},
  {Bridle}, {Clowe}, {Dahle}, {Ellis}, {Erben}, {Hetterscheidt}, {High},
  {Hirata} \& [...]}{{Massey} et~al.}{2007}]{massey07}
{Massey} R.,  {Heymans} C.,  {Berg{\'e}} J.,  {Bernstein} G.,  {Bridle} S.,
  {Clowe} D.,  {Dahle} H.,  {Ellis} R.,  {Erben} T.,  {Hetterscheidt} M.,
  {High} F.~W.,  {Hirata} C.,    [...] 2007, \mnras, 376, 13

\bibitem[\protect\citeauthoryear{{Massey}, {Rhodes}, {Leauthaud}, {Ellis},
  {Scoville} \& {Finoguenov}}{{Massey} et~al.}{2006}]{massey06}
{Massey} R.,  {Rhodes} J.,  {Leauthaud} A.,  {Ellis} R.,  {Scoville} N.,
  {Finoguenov} A.,  2006, in American Astronomical Society Meeting Abstracts
  Vol.~38 of Bulletin of the American Astronomical Society, {A Direct View of
  the Large-Scale Distribution of Mass, from Weak Gravitational Lensing in the
  HST COSMOS Survey}.
pp 966--+

\bibitem[\protect\citeauthoryear{{Massey}, {Stoughton}, {Leauthaud}, {Rhodes},
  {Koekemoer}, {Ellis} \& {Shaghoulian}}{{Massey} et~al.}{2010}]{massey10}
{Massey} R.,  {Stoughton} C.,  {Leauthaud} A.,  {Rhodes} J.,  {Koekemoer} A.,
  {Ellis} R.,    {Shaghoulian} E.,  2010, \mnras, 401, 371

\bibitem[\protect\citeauthoryear{{Mead}, {King} \& {McCarthy}}{{Mead}
  et~al.}{2010}]{mead10}
{Mead} J.~M.~G.,  {King} L.~J.,    {McCarthy} I.~G.,  2010, \mnras, 401, 2257

\bibitem[\protect\citeauthoryear{{Meneghetti}, {Fedeli}, {Zitrin},
  {Bartelmann}, {Broadhurst}, {Gottl{\"o}ber}, {Moscardini} \&
  {Yepes}}{{Meneghetti} et~al.}{2011}]{meneghetti11}
{Meneghetti} M.,  {Fedeli} C.,  {Zitrin} A.,  {Bartelmann} M.,  {Broadhurst}
  T.,  {Gottl{\"o}ber} S.,  {Moscardini} L.,    {Yepes} G.,  2011, \aap, 530,
  A17+

\bibitem[\protect\citeauthoryear{{Miyazaki}, {Komiyama}, {Sekiguchi},
  {Okamura}, {Doi}, {Furusawa}, {Hamabe}, {Imi}, {Kimura}, {Nakata}, {Okada},
  {Ouchi}, {Shimasaku}, {Yagi} \& {Yasuda}}{{Miyazaki}
  et~al.}{2002}]{miyazaki02}
{Miyazaki} S.,  {Komiyama} Y.,  {Sekiguchi} M.,  {Okamura} S.,  {Doi} M.,
  {Furusawa} H.,  {Hamabe} M.,  {Imi} K.,  {Kimura} M.,  {Nakata} F.,  {Okada}
  N.,  {Ouchi} M.,  {Shimasaku} K.,  {Yagi} M.,    {Yasuda} N.,  2002, \pasj,
  54, 833

\bibitem[\protect\citeauthoryear{{Moody}, {Turner} \& {Gott} III}{{Moody}
  et~al.}{1983}]{moody83}
{Moody} J.~E.,  {Turner} E.~L.,    {Gott} III J.~R.,  1983, \apj, 273, 16

\bibitem[\protect\citeauthoryear{{Novikov}, {Colombi} \& {Dor{\'e}}}{{Novikov}
  et~al.}{2006}]{novikov06}
{Novikov} D.,  {Colombi} S.,    {Dor{\'e}} O.,  2006, \mnras, 366, 1201

\bibitem[\protect\citeauthoryear{{Oguri} \& {Hamana}}{{Oguri} \&
  {Hamana}}{2011}]{OH11}
{Oguri} M.,  {Hamana} T.,  2011, \mnras, 414, 1851

\bibitem[\protect\citeauthoryear{{Pimbblet} \& {Drinkwater}}{{Pimbblet} \&
  {Drinkwater}}{2004}]{PD04}
{Pimbblet} K.~A.,  {Drinkwater} M.~J.,  2004, \mnras, 347, 137

\bibitem[\protect\citeauthoryear{{Rasmussen}}{{Rasmussen}}{2007}]{rasmussen07}
{Rasmussen} J.,  2007, ArXiv e-prints

\bibitem[\protect\citeauthoryear{{Rhodes}, {Refregier} \& {Groth}}{{Rhodes}
  et~al.}{2000}]{rhodes00}
{Rhodes} J.,  {Refregier} A.,    {Groth} E.~J.,  2000, \apj, 536, 79

\bibitem[\protect\citeauthoryear{{Rhodes}, {Massey}, {Albert}, {Collins},
  {Ellis}, {Heymans}, {Gardner}, {Kneib}, {Koekemoer}, {Leauthaud}, {Mellier},
  {Refregier}, {Taylor} \& {Van Waerbeke}}{{Rhodes} et~al.}{2007}]{rhodes07}
{Rhodes} J.~D.,  {Massey} R.~J.,  {Albert} J.,  {Collins} N.,  {Ellis} R.~S.,
  {Heymans} C.,  {Gardner} J.~P.,  {Kneib} J.-P.,  {Koekemoer} A.,  {Leauthaud}
  A.,  {Mellier} Y.,  {Refregier} A.,  {Taylor} J.~E.,    {Van Waerbeke} L.,
  2007, \apjs, 172, 203

\bibitem[\protect\citeauthoryear{{Rix}, {Barden}, {Beckwith}, {Bell}, {Borch},
  {Caldwell}, {H{\"a}ussler}, {Jahnke}, {Jogee}, {McIntosh}, {Meisenheimer},
  {Peng}, {Sanchez}, {Somerville}, {Wisotzki} \& {Wolf}}{{Rix}
  et~al.}{2004}]{rix04}
{Rix} H.-W.,  {Barden} M.,  {Beckwith} S.~V.~W.,  {Bell} E.~F.,  {Borch} A.,
  {Caldwell} J.~A.~R.,  {H{\"a}ussler} B.,  {Jahnke} K.,  {Jogee} S.,
  {McIntosh} D.~H.,  {Meisenheimer} K.,  {Peng} C.~Y.,  {Sanchez} S.~F.,
  {Somerville} R.~S.,  {Wisotzki} L.,    {Wolf} C.,  2004, \apjs, 152, 163

\bibitem[\protect\citeauthoryear{{Schneider}, {King} \& {Erben}}{{Schneider}
  et~al.}{2000}]{schneider00}
{Schneider} P.,  {King} L.,    {Erben} T.,  2000, \aap, 353, 41

\bibitem[\protect\citeauthoryear{{Seitz} \& {Schneider}}{{Seitz} \&
  {Schneider}}{1995}]{seitz95}
{Seitz} C.,  {Schneider} P.,  1995, \aap, 297, 287

\bibitem[\protect\citeauthoryear{{Sheth} \& {Jain}}{{Sheth} \&
  {Jain}}{2003}]{SJ03}
{Sheth} R.~K.,  {Jain} B.,  2003, \mnras, 345, 529

\bibitem[\protect\citeauthoryear{{Skilling}}{{Skilling}}{1998}]{skilling98}
{Skilling} J.,  1998, in {G.~J.~Erickson, J.~T.~Rychert, \& C.~R.~Smith} ed.,
  Maximum Entropy and Bayesian Methods {Massive Inference and Maximum Entropy}.
pp~1--+

\bibitem[\protect\citeauthoryear{{Sousbie}, {Pichon}, {Courtois}, {Colombi} \&
  {Novikov}}{{Sousbie} et~al.}{2006}]{sousbie06a}
{Sousbie} T.,  {Pichon} C.,  {Courtois} H.,  {Colombi} S.,    {Novikov} D.,
  2006, ArXiv Astrophysics e-prints

\bibitem[\protect\citeauthoryear{{Williams}, {Mathur}, {Nicastro} \&
  {Elvis}}{{Williams} et~al.}{2007}]{williams07}
{Williams} R.~J.,  {Mathur} S.,  {Nicastro} F.,    {Elvis} M.,  2007, \apj,
  665, 247

\bibitem[\protect\citeauthoryear{{Williams}, {Mulchaey}, {Kollmeier} \&
  {Cox}}{{Williams} et~al.}{2010}]{williams10}
{Williams} R.~J.,  {Mulchaey} J.~S.,  {Kollmeier} J.~A.,    {Cox} T.~J.,  2010,
  \apjl, 724, L25

\bibitem[\protect\citeauthoryear{{Wuyts}, {van Dokkum}, {Kelson}, {Franx} \&
  {Illingworth}}{{Wuyts} et~al.}{2004}]{wuyts04}
{Wuyts} S.,  {van Dokkum} P.~G.,  {Kelson} D.~D.,  {Franx} M.,    {Illingworth}
  G.~D.,  2004, \apj, 605, 677

\bibitem[\protect\citeauthoryear{{Yess} \& {Shandarin}}{{Yess} \&
  {Shandarin}}{1996}]{YS96}
{Yess} C.,  {Shandarin} S.~F.,  1996, \apj, 465, 2

\bibitem[\protect\citeauthoryear{{York}, {Adelman}, {Anderson} Jr., {Anderson},
  {Annis}, {Bahcall}, {Bakken}, {Barkhouser}, {Bastian}, {Berman}, {Boroski} \&
  [...]}{{York} et~al.}{2000}]{york00}
{York} D.~G.,  {Adelman} J.,  {Anderson} Jr. J.~E.,  {Anderson} S.~F.,  {Annis}
  J.,  {Bahcall} N.~A.,  {Bakken} J.~A.,  {Barkhouser} R.,  {Bastian} S.,
  {Berman} E.,  {Boroski} W.~N.,    [...] 2000, \aj, 120, 1579

\bibitem[\protect\citeauthoryear{{Zitrin}, {Broadhurst}, {Rephaeli} \&
  {Sadeh}}{{Zitrin} et~al.}{2009}]{zitrin09b}
{Zitrin} A.,  {Broadhurst} T.,  {Rephaeli} Y.,    {Sadeh} S.,  2009, \apjl,
  707, L102

\end{thebibliography}

\label{lastpage}

\end{document}